\documentclass[11pt]{article}
\usepackage{latexsym} %\usepackage{epsf}
\usepackage{amsfonts}
\usepackage{amsmath}
\usepackage{amsthm}
\usepackage{amssymb}
\renewcommand{\theequation}{\thesection.\arabic{equation}}
\newcounter{subequation}[equation]
\makeatletter

\newcommand{\p}{^{\prime}}

\newcommand{\lb}{\label}

\expandafter\let\expandafter\reset@font\csname reset@font\endcsname

\def\subeqnarray{\arraycolsep1pt
    \def\@eqnnum\stepcounter##1{\stepcounter{subequation}%
        {\reset@font\rm(\theequation\alph{subequation})}}
\jot5mm     \eqnarray}

\makeatother

\def\be{\begin{equation}}
\def\ee{\end{equation}}
\def\bea{\begin{eqnarray}}
\def\eea{\end{eqnarray}}
\def\ba{\begin{array}}
\def\ea{\end{array}}
\def\dd{\partial}

\def\half{\frac{1}{2}}

\def\one#1{#1^{\raise5pt\hbox{$\scriptstyle\!\!\!\!1$}}\,{}}
\def\two#1{#1^{\raise5pt\hbox{$\scriptstyle\!\!\!\!2$}}\,{}}

\def\tilde{\widetilde}
\def\II{\hbox{{1}\kern-.25em\hbox{l}}}
\def\a{\alpha}
\def\b{\beta} 
\def\c{\gamma}
\def\d{\delta}
\def\e{\varepsilon}

\makeatletter
\def\binrel@#1{\begingroup
  \setboxz@h{\thinmuskip0mu
    \medmuskip\m@ne mu\thickmuskip\@ne mu
    \setbox\tw@\hbox{$#1\m@th$}\kern-\wd\tw@
    ${}#1{}\m@th$}%
  \edef\@tempa{\endgroup\let\noexpand\binrel@@
    \ifdim\wdz@<\z@ \mathbin
    \else\ifdim\wdz@>\z@ \mathrel
    \else \relax\fi\fi}%
  \@tempa
}
\let\binrel@@\relax
\def\overset#1#2{\binrel@{#2}%
  \binrel@@{\mathop{\kern\z@#2}\limits^{#1}}}
\def\underset#1#2{\binrel@{#2}%
  \binrel@@{\mathop{\kern\z@#2}\limits_{#1}}}
\makeatother
\newfont{\bbd}{msbm10 scaled\magstep1}

\newtheorem{prop}{Proposition}

\begin{document}

\begin{center}

     {\LARGE {
Representations of  orthogonal and symplectic Yangians\footnote{
\large \sl Dedicated to Harald Sorg. } } }

\vspace{.3cm}

{\large \sf
D. Karakhanyan$^{a}$\footnote{\sc e-mail: karakhan@yerphi.am  }, 
R. Kirschner$^b$\footnote{\sc e-mail:Roland.Kirschner@itp.uni-leipzig.de}} 

%\vspace{0.3cm}

\begin{itemize}
\item[$^a$]
{\it Yerevan Physics Institute,
2 Alikhanyan br., 0036 Yerevan, Armenia }
\item[$^b$]
{\it Institut f\"ur Theoretische
Physik, Universit\"at Leipzig, \\
PF 100 920, D-04009 Leipzig, Germany}
\end{itemize}
\end{center}

\vspace{.3cm}
\begin{abstract}
\noindent
Extended Yangian algebras of orthogonal and symplectic types are defined by
the Yang-Baxter RLL relation involving the fundamental R-matrix with 
$so(n)$ or $sp(2m)$ symmetry. We study representations of highest weight
characterized by weight function ratios. We consider the algebra relations
for the linear and the quadratic evaluations and the resulting conditions
imposed on the representation weights. We present expressions of 
L-operators constructed on underlying Clifford and Heisenberg algebras and
characterize their representations. 
\end{abstract}

%\vspace{.5 cm}

    %%%%%%%%%%%%%%%%%%%%%%%%%%%%%%%%%%%%%%%%%%%%%%%%%%%%%%%%%%%%%%%%%%%%%%%%%%%%%%
 %   {\small \tableofcontents}
    \renewcommand{\refname}{References.}
    \renewcommand{\thefootnote}{\arabic{footnote}}
    \setcounter{footnote}{0}

%\newpage

\section{Introduction}

The treatment of integrable quantum systems is based on the Yang-Baxter
equations \cite{FST,TTF,KuSk1,Fad,ZZ,Drinfeld,Drin}. The cases of symmetry
with respect to the Lie algebras $\mathcal{G} = sp(2m), so(2m), so(2m+1)$ 
are known to be more involved compared to the cases $\mathcal{G}
=g\ell(n)$. The fundamental $R$ matrix, on which the formulation of the
Yangian algebra can be based, is linear in the spectral parameter in the
$g\ell(n)$ cases but quadratic in the former ones \cite{ZZ,BKWK}.
 In the $g\ell(n)$  case linear evaluation 
$L$ operators exist without restrictions on the Lie algebra representations,
whereas in the orthogonal and symplectic cases linear $L$ exist in distinguished
cases only \cite{Re}. 
Monodromy matrices are $L$ operators composed by products of simple $L$
operators and play a central role in the treatment of quantum spin chains.
Simple $L$ operators are of interest as factors in the monodromy matrices.

The Yang-Baxter formulation of Yangians of the orthogonal and symplectic
types has been studied in \cite{AAC03,AMR05}. Independent proofs of the  equivalence 
of this formulation to the formulations by Drinfeld \cite{Drinfeld,Drin}
have been given in \cite{JLM17} and \cite{GRW17}. 
The successful applications of results and methods of integrable quantum
systems to questions of gauge field theories, e.g.
\cite{LevPadua,BS03,DHP09,CDK13}, motivated recent
studies of Yangians with symmetries related to conformal transformations in  
Quantum field theory \cite{CDI1,IKK15,FIKK,KK17}.

The similarities of the orthogonal and symplectic Lie algebras can be traced
back to the existence of an invariant metric in the spaces of their
fundamental representations. The similarities extend to the Yangians of the
corresponding types and allow a uniform treatment of the Yang-Baxter
relations as proposed e.g. in
\cite{IKK15} and generalized to the ortho-symplectic super-algebra case in
\cite{FIKK}. 

Representations of the Yangians of orthogonal and symplectic types have been
studied in \cite{AMR05}. Examples of $L$ operators linear or quadratic in
the spectral parameter have been considered in \cite{IKK15}. The algebra
relations of the corresponding Yangians with the expansion of the $L$ operators    
truncated at the second non-trivial term have been formulated in
\cite{KK17}. 

In the present paper we consider representations of Yangian
algebras the generators of which arise from $L$ operators linear or quadratic 
in the spectral parameter. We derive the conditions on the representation
weights implied by the algebra relations of the second order evaluation
of Yangians of the orthogonal and symplectic types. In this way we
describe the situations of simple $L$ operators. 
We give explicit constructions of $L$ operators based on underlying algebras
of Clifford and Heisenberg type. 
Some results appeared in the conference contributions \cite{KKc19}.

In Section 2 we formulate the Yang-Baxter equations used to define the
extended Yangian algebra of orthogonal and symplectic types in terms of the
expansion of $L$ operators. We formulate also the cases of linear and quadratic
evaluations and the highest weight conditions.

In Section 3 we summarize the results of the representation weights for linear and
quadratic evaluations and of explicit $L$ operators before entering the
detailed discussions of the following sections.

In Section 4 we investigate the conditions of highest weight and consider the
relations of the weight functions as derived from the center generating
function. We define the simple weight function ratios, associate with them
$g\ell(2)$ type Yangian algebra representations and consider the conditions for
finite-dimensional representations. 

In Section 5 the algebra relations of the quadratic evaluation are given.
The resulting  conditions on the representation weights are studied in
detail. Examples of admitted weights and weight function ratios are
presented.

We construct explicit linear $L$ operators in Section 6 and quadratic
in Section 7. 

The results are discussed in Section 8.

\section{Orthogonal and symplectic Yangians}
\setcounter{equation}{0}

%\subsection{Extended Yangian algebra}

The formulation of the Yangian algebra of type $\mathcal{G}$ can be based on
the fundamental Yang-Baxter $R$ matrix. In the case of symmetry with respect
to the Lie algebra $\mathcal{G} = so(n) $ $n=2m$ or  $ n=2m+1$ or \ 
$\mathcal{G}= sp(n)$, $n=2m$, 
it has the form \cite{ZZ,BKWK,Re,API}
   \be\label{rzW}
    R_{b_1b_2}^{a_1a_2}(u)=u(u + \beta)
    I^{a_1a_2}_{b_1b_2}
    +(u + \beta)P^{a_1a_2}_{b_1b_2}
    - \epsilon \, u \, K^{a_1a_2}_{b_1b_2} \; ,
    \ee
    where 
\be\label{IKP}
I^{a_1a_2}_{b_1b_2} = \delta^{a_1}_{b_1} \delta^{a_2}_{b_2} \; , \qquad
P^{a_1a_2}_{b_1b_2} = \delta^{a_1}_{b_2} \delta^{a_2}_{b_1} \; , \qquad
K^{a_1a_2}_{b_1b_2} = \varepsilon^{a_1a_2} \, \varepsilon_{b_1b_2}  \; ,
\qquad    \beta = \frac{n}{2} -\epsilon.
    \ee
The
sign factor $\epsilon$ distinguishes the orthogonal ($\epsilon=+1$) and 
the symplectic ($\epsilon= -1$) cases.

Here $\varepsilon_{ab}$ is a non-degenerate
 invariant metric in the  space  $V_f$ of the fundamental representation.
It has the $\e$ symmetry property
\be \label{syme}
 \e_{ab} = \epsilon \e_{ba}, 
 \ee
and allows to lower and rise indices by the rules (sum over repeated
indices)
$$ 
x_a=\e_{ab} x^b,\qquad x^a=\e^{ab}x_b,\qquad\e_{ac}\e^{cb} =\delta_a^b.
$$

The generators $ (L^{(k)})^a_b $ of the extended Yangian algebra  
$\mathcal{Y}(\mathcal{G})$ appear in the expansion of the $L$ operator
\be
 \label{defYan}
 L^a_b(u) = \sum_{k=0}^\infty  \frac{(L^{(k)})^a_b}{u^k}  \; , \qquad\qquad
 L^{(0)} = I \; ,
 \ee
which satisfies  the Yang-Baxter RLL-relations
 \be
 \label{rll}
 \begin{array}{c}
    R^{a_1a_2}_{b_1b_2}(u-v)L^{b_1}_{c_1}(u)L^{b_2}_{c_2}(v)=
    L^{a_2}_{b_2}(v)L^{a_1}_{b_1}(u)R^{b_1b_2}_{c_1c_2}(u-v). 
  \end{array}
    \ee
Here $L(u)$ is an algebra valued matrix depending on the spectral 
parameter $u$ and plays the role of the generating function of the 
extended Yangian algebra generators, $L \in (End \ V_f) \otimes \mathcal{Y}$. 
Related to the co-multiplication and the applications to quantum spin 
chains it is also called (matrix of) monodromy.

The algebra  $ \mathcal{Y}$ defined by (\ref{rll}), (\ref{rzW}) does not
depend on the choice of the non-degenerate metric with the $\epsilon$-
symmetry (\ref{syme}). The metric may be specified in particular studies 
for convenience. For descending to particular real forms of the Lie algebra 
particular metric signatures may be preferable.

Factorizing the central elements leads to the 
Yangian algebra definition equivalent to the original  definitions by
Drinfeld \cite{Drinfeld,Drin}. Independent  proofs of the equivalence are given in
\cite{JLM17} and  \cite{GRW17}.

We write the RLL relation (\ref{rll}) in the commutator form  lowering all
indices,
\be \label{LuLvosp1}
 (u-v) (u-v+\beta) [L_{a_1, b_1}(u), L_{a_2, b_2}(v) ] = \ee
$$  (u-v+\beta)
\Big[ L_{a_2, b_1}(v)L_{a_1, b_2}(u) -  
L_{a_2, b_1}(u) L_{a_1, b_2}(v)
\Big ] + $$ $$ (u-v)  \Big [\epsilon \e_{a_1 a_2} (L^t(u)L(v))_{b_1,b_2}   
-  \e_{b_1 b_2}(L(v)L^t(u))_{a_2, a_1} \Big ].
$$

The algebra relations in (\ref{LuLvosp1}) 
involving the first term in the expansion
(\ref{defYan}) only, which will be denoted in the following by $G, L^{(1)} = G$, 
are easily seen to coincide with the commutation relations characterizing the
underlying Lie algebra $\mathcal{G}$,

\be \label{Liealg0}
[ G_{ab},  G_{cd} ]=-\e_{cb}  G_{ad} + \e_{ad} G_{cb} + 
\e_{ac} G_{bd}-\e_{db} G_{ca}.
\ee

Further, the  generators appearing in the expansion (\ref{defYan})
at higher order $k>1$ obey the relations of the adjoint action,

\be \label{Lieadj}
[ G_{ab},  L^{(k)}_{c,d} ]=-\e_{cb}  L^{(k)}_{a,d} + \e_{ad} L^{(k)}_{c,b} + 
\e_{ac} L^{(k)}_{b,d}-\e_{d,b} L^{(k)}_{c,a}.
\ee
The center of the extended Yangian algebra of
the orthogonal and symplectic types
has been analysed in  \cite{AMR05} and formulated in terms of the generating
function $C(u)$. 
 defined as
\be\label{c}
C(u)=L^t(u-\b)L(u)=c(u)I.
\ee

The statement that $C(u)$ is central implies that the  
$\epsilon$ symmetric part of
$L^{(p)} $, $L_{ab}^{(p)} + \e L_{ba}^{(p)}$,   
is expressed in terms of products of
lower degree generators and central elements. The  $\epsilon$ antisymmetric part of $L^{(p)}$
can be considered as the independent generators of the Yangian algebra.
A proof will be given below ...

We choose the index range
$ a,b,= -m, ...,-1,(0), +1, ..., +m$, where the index value $0$ appears in the
$so(2m+1)$ case only, and denote the indices also by $\pm i, \pm j,
i,j = 1, ...,m$ (and $0$ in the $so(2m+1)$ case  ).

%%%%%%%%%%%%%%%%%%%%%%%%%%%%%%%%%%%%%%%%%%%%%%%%%%%%%%%%%%%%%%%%%%%%%%%%
In the following we compare also with the extended 
Yangian of the $g\ell(n)$ type, defined by the RLL relation of the
form (\ref{rll}) with the $R$ matrix 
 substituted as $R=I + P$, where the
matrix elements of $I$ and $P$ are as in (\ref{IKP}).
We shall use the notation $L^{g\ell}(u)$ for the corresponding $L$ operators.
The resulting commutation relation of the latter differs in form from 
(\ref{LuLvosp1}) by deleting the terms of the last line. 

\be \label{rllgln}
 (u-v) [L^{g\ell}_{a_1, b_1}, L^{g\ell}_{a_2, b_2} ] = L^{g\ell}_{a_2, b_1}(v) 
L^{g\ell}_{a_1, b_2}(u) - 
L^{g\ell}_{a_2, b_1}(u) L^{g\ell}_{a_1, b_2}(v). 
\ee

 We specify
the metric as
\be \label{metric}
\e_{ab} = \e_a \delta_{a, -b}, \qquad\qquad \e_i = 1,\qquad\qquad \e_{-i} = \epsilon, 
\ee

Comparing (\ref{rllgln}) and (\ref{LuLvosp1}), 
we find $\mathcal{Y}(g\ell(m+1))$  algebras in
$\mathcal{Y}(\mathcal{G}), \mathcal{G}= so(2m), $ $ sp(2m), so(2m+1)$.
$\mathcal{Y}^{\sigma_a, \sigma_b}$ are generated by the subsets
of $L_{a,b}$ with all indices $a$ of the same sign $\sigma_a$ and all indices
$b$ also of the same sign $\sigma_b$ and no index value equal to $0$ (for the
$so(2m+1)$ case). Also the co-product is modified by the restriction to the
corresponding index range.

The conditions on the vector of the highest weight $|0\rangle$  are ($i,j =1,
...,m$) 
\be \label{hwY}
 L_{-i,j}(u) |0\rangle = 0 ,\quad i<j, \qquad L_{-i,-j}(u) |0\rangle = 0 ,\qquad
  L_{-i,i}(u) |0\rangle =\lambda_i(-u) |0\rangle, 
 \ee
In the case of $so(2m+1)$
we add the condition
\be \label{hwY0}
 L_{-i,0}(u) |0\rangle = 0. 
 \ee
 The statement about the dependence of the $\epsilon$ symmetric parts of 
 the generators implies further
\be \label{hwYadd}
 L_{j, -i}(u) |0\rangle = 0,\quad i>j,\qquad L_{i,-i}(u) |0\rangle =\lambda_{-i}
 (-u) |0\rangle, 
\ee
$$
L_{0,-i}(u)|0\rangle = 0,\qquad\qquad L_{0,0}(u) |0\rangle = \lambda_0(-u) |
0\rangle, 
$$
where these additional weight functions are dependent on the components of
the basic weight functions $\lambda_i(-u)$. This dependence will be written
explicitly in the case of quadratic evaluation below.

We may rewrite the highest weight conditions in terms of $L^a_{ \ b}(u)$.
Then the weight functions appear as eigenvalues of the diagonal matrix
elements. By a similarity transformation a formulation with a matrix
$\tilde L(u) $ can be found, such that the elements annihilating 
$|0>$ are all above or all below the diagonal.

We formulate the highest weight condition for the case of $g\ell(n)$ as
($\alpha, \beta = 1, ..., n$)
\be \label{hwYgl}
L^{g\ell}_{\alpha \beta}(u)|0\rangle=0,\qquad \alpha<\beta,\qquad
L^{g\ell}_{\alpha \alpha}(u)|0\rangle=\lambda_{\alpha}^{g\ell}(-u)|0\rangle.
\ee

The mentioned statement about the dependence of the $\e$-symmetric part of
the generators $L^{(p)} $
 implies that we have only $m$ independent weight functions, e.g.
$\lambda_i(u), i= 1, ..., m $. Moreover,  equivalent representations 
appear with sets of weight functions differing by a common factor,
$$
\tilde \lambda_a(u) = \phi(u) \lambda_a(u), \qquad\qquad
a=-m, ..,-1, (0), +1, ..., m. 
$$
To characterize a representation one should give $m$ different
weight function ratios.

The set of $m$ ratios 
including the $m-1$
\be \label{fi}
 \frac{\lambda_i(u)}{\lambda_{i+1}(u)} = f_i(u),\qquad\qquad i=1, ...,m-1, 
 \ee
and, specific for the particular cases,
$$
 \frac{\lambda_m(u)}{\lambda_{1-m}(u) } = f_m(u),\qquad  {\rm for} \qquad
so(2m), 
$$ 
$$
- \frac{\lambda_m(u)}{\lambda_{-m}(u) } = f_m(u), \qquad  {\rm for}\qquad
sp(2m), 
$$
$$
 \frac{\lambda_m(u)}{\lambda_{0}(u) } = f_m(u) ,\qquad  {\rm for} \qquad
so(2m+1), 
$$
is chosen  related to the simple roots in our form of the highest weight conditions.
We are interested in the case of finite order evaluation, where the
expansion in (\ref{defYan}) is truncated, $L^{(k)}_{ab} = 0, \ k>s$. In this
case
$f_i(u)$ are ratios of polynomials and thus they are given by $s$
$m$-tuples of  zeros and poles
$$u^+_{i,k},\;\; u_{i,k} , \qquad i=1, ...,m; \qquad k=1, ..,s.  $$

Each of the basic  weight function ratios of the considered Yangian representations 
can be associated with a $g\ell(2)$ type Yangian representation. This will
be discussed in detail in Section 4.

In this paper we study representations of linear 
and quadratic evaluations of Yangian algebras of the orthogonal and
symplectic types. In the quadratic evaluation case the expansion of $L(u)$
(\ref{defYan} ) 
is truncated after the second non-trivial term, $L^{(k)}_{ab} = 0$ for
$k>2$.  
Multiplying $L(u)$ by $u^2$ we put
\be \label{LGH}
 L(u) = u^2 I + u G + H, \ee
changing the notations as $L^{(1)} = G $ and $L^{(2)} = H$. 
In the case of linear evaluation (\ref{defYan} ) is truncated after the frist
non-trivial term. Multiplying $L(u)$ by $u$ we put is this case
\be \label{LG}
 L(u) = u I + G. \ee
The algebra relations have been formulated in \cite{KK17} and their
connection to the center generating function $C(u)$ will be explained below 
in Section 5. 
  
In the quadratic case the weight functions can be writtes as
\be \label{lambu2}
\lambda_a(-u) = u^2 I + u \lambda_a^{[1]} +  \lambda_a^{[2]}. \ee 
We decompose the last contribution
$ \lambda_a^{[2]} = \tilde \lambda_a^{[2]} + \bar \lambda_a^{[2]}, $
where 
$ \tilde \lambda_a^{[2]} = \e \tilde \lambda_{-a}^{[2]}, \bar \lambda_a^{[2]}
= -\e \bar \lambda_{-a}^{[2]}. $
The algebra relations imply
$$ \lambda_i^{[1]} = -\e  \lambda_{-i}^{[1]}, \ \   \lambda_0^{[1]} = 0.  $$
$ \lambda_i^{[1]}$ is the eigenvalue of $G_{-i i}$ in action on the highest
weight vector $|0>$. 
$\tilde \lambda_a^{[2]} $ depends on the components $\tilde \lambda_a^{[1]}
$, it turns out to be the eigenvalue of $\half (G^2 + \beta G + k I)_{-a,a}$ in
action on $|0>$. $k$ denotes  a central element. 

\section{Results on representations and $L$ operators }
\setcounter{equation}{0}

We present a summary of the results on the representation weights
allowed by the conditions derived from the 
linear and quadratic evaluation Yangian algebra relations.
Detailed discussions and proofs will be given in the next sections.
We present also examples of explicit $L$ operators and weights of 
corresponding representations.  These examples do not cover all cases of 
solutions of the weight conditions. However they include cases with
the maximal allowed number of continuous parameters in the weights.

The conditions on the weight functions can be formulated in terms of
\be \label{Lambi}
 \Lambda_i(u,\alpha,\gamma) = \lambda_i(-u+\alpha) \lambda_{-i}(-u+\gamma)
\ee
as the $m-1$ conditions, $i=1, .., m-1$,
\be \label{Lii+1}
 \Lambda_i(u, \beta -i+1, 1) = \Lambda_{i+1}(u,\beta-i+1, 1) \ee
and, in the case of $so(2m+1)$ the additional condition
\be \label{Lm0}
 \Lambda_m(u, \half, 1) = \Lambda_0(u, \half, 1). \ee 

These conditions hold in general for  representations built on the highest
weight vector $|0\rangle $ obeying (\ref{hwY}) (\ref{hwY0}). They are derived from the
action of the center generating function $C(u)$ on $|0\rangle $.
They correspond to the
statement of the Proposition 5.14 in \cite{AMR05}.

\subsection{The case of linear evaluation}
  
From [\ref{LG}) and (\ref{lambu2}) we see that
 the weight functions are to be substituted as 
\be \label{lamblin1}
 \lambda_i(-u) = u \e + \lambda_i, \ee
with $\lambda_a^{[2]} = 0$,  $\lambda_i^{[1]}= \lambda_i$,  
and obtain the $m-1$ conditions (\ref{Lii+1}) in the two-factor form
\be \label{lambdalins}
(\lambda_{i+1} - \lambda_i) \big(\lambda_{i+1} +
\lambda_i- \epsilon (\beta-i)\big) = 0. 
\ee
In the case $so(2m+1)$ the additional relation (\ref{Lm0}) reduces 
in the linear evaluation case with $\lambda_0^{[1]}= 0,$  $\epsilon =+1 $ to
\be \label{lambdalin0s}
 \lambda_m \big( \lambda_m + \half\big) = 0 .
\ee

The case of all weights $\lambda_i, i=1, ...,m$, being equal is a solution
of the $m-1$ conditions, 
\be \label{lamb1m}
 (\lambda_1, ..., \lambda_m) = (\lambda, ..., \lambda). \ee
In the cases of $so(2m)$ and $sp(2m)$ the parameter $\lambda$ can take any
value.
In the case of $so(2m+1)$ the additional condition resticts the value of
$\lambda $ to $0$ or $- \half$. The sequence of weight function ratios
(\ref{fi}),
$f_1(u), ...,f_{m-1}, f_m(u)$ has in the equal weight examples the form
$1, ...,1, f_m(u)$ with $f_m(u) = \frac{u-\e \lambda}{u+\e \lambda} $ in the
cases of  $so(2m)$ and $sp(2m)$ and $f_m(u) =1 $ or $ \frac{u+\half}{u} $  in 
the case of $so(2m+1)$. 

The general solution for the sequence of weights can be described as
consisting of segments of equal values, e.g.
\be \label{lambdamus} 
 ...., \lambda, ..., \lambda, \mu, ..., \mu, ..., \ee
or in other terms, $ \lambda_{i_1+1} = \lambda_{i_1+2}= ...= \lambda_{i_2} = \lambda,
\lambda_{i_2+1} = \lambda_{i_2+2}= ...=\lambda_{i_3} =\mu $,
where the two values  are related by the condition of vanishing of the
second bracket in (\ref{lambdalins} ) at $i = i_2$, 
$ \mu = - \lambda + \e(\beta-i_2) $.

In the cases of $so(2m)$ and $sp(2m)$ there is one free parameter in the
solutions.
In the case of $so(2m+1)$ the additional condition (\ref{lambdalin0s}) restricts the value 
in the last segment containing
$\lambda_m $ to $0$ or $- \half$. Thus in this case there is no free parameter in the
solutions. 

We consider two constructions of $L$ operators  $L(u) = Iu + G$. The first
construction is well known \cite{Re} and has been considered in many papers. 
The second example appeared recently in \cite{RF20} for the case $so(2m)$ with a
different way of construction.

\vspace{.5cm}

{\it 1. Presentations on Clifford algebras }

The $L$ operator can be constructed
explicitly in terms of the underlying quantum oscillator or Clifford
algebra, generated by $c^a, a=-m, ...,(0), ...+m$, obeying
   
 \be\label{osc1}
c^ac^b+\epsilon c^bc^a=\e^{ba}
 \ee
as
$$L_{ab}(u) = u \e_{ab} - \half (c_a c_b - c_b c_a) . $$
A Yangian representation of the types $so(2m)$ or $sp(2m)$ of highest weight 
can be built on the vector 
$|0\rangle$ obeying 
$$c_{-i} |0\rangle = 0, \ i=1, ...,m $$
or on the vector $|\tilde 0 \rangle> = c_m |0\rangle$. In the case
$so(2m+1)$ the second opportunity is excluded.

Of course, the resulting sequences of weights are solutions of the
two-factor
conditions (\ref{lambdalins}), (\ref{lambdalin0s}).  However here is no free parameter 
involved.

\vspace{.5cm}

{\it 2. Presentations on Heisenberg algebras}

This example results in representations of the linear evaluation of the
Yangians of types $so(2m), sp(2m)$
depending on one parameter.
The matrix of generators $G^a_b$ is constructed in $m \times m$ block form
$$  G = 
\begin{pmatrix}
\hat I & 0 \\
-\hat x & \hat I
\end{pmatrix}
\begin{pmatrix}
- \ell \hat I -\beta \hat I & \hat \dd \\
0 & \ell
\end{pmatrix}
\begin{pmatrix}
\hat I & 0 \\
\hat x & \hat I
\end{pmatrix}.
$$
where $\hat I$ is the unit matrix and $\hat x$ and $\hat \dd$ are
$\e$-antisymmetric matrices the entries of which obey 
\be \label{ddijxkls}
[\dd^i_j, x^k_l] = \delta^k_j \delta^i_l - \epsilon \delta^i_k \delta^j_l. \ee
This construction involves $\ell$ as a parameter. It applies for $so(2m)$
and $sp(2m)$ and does not work for $so(2m+1)$. 
The representation is spanned by polynomials in $x^i_k$ and the role of the highest
weight vector is played by the constant function $1$. 
We obtain a realization of the particular solution (\ref{lamb1m}) with
$\lambda = -\e \ell $. 

\subsection{The case of quadratic evaluation}

We consider the  conditions on the weight components resulting from 
(\ref{Lii+1}) in the case
of the quadratic evaluation, where the weight functions are to be
substituted as (\ref{lambu2}). They depend on two components,
$\lambda^{[1]}_i = \lambda_i$ and $\bar \lambda^{[2]}_i = \bar \lambda_i$,
$i=1, ...,m$. A particular solution is the case of equal values,
$$ (\lambda_1, ..., \lambda_m) =
(\lambda, ..., \lambda); \ \ ( \bar\lambda_1, ..., \bar \lambda_m) = (\bar\lambda, ..., \bar\lambda). $$
In the cases of $so(2m)$ and $sp(2m)$ the parameters $\lambda$ and $\bar
\lambda $ can take any value.
In the case of $so(2m+1)$ the additional condition (\ref{Lm0})
restricts the values and only one parameter is free. 

In general, $\lambda_i $ and $\bar \lambda_j$ are related by the $m-1$
conditions
\be \label{u1is}
\bar \lambda_i \big(\lambda_i - \epsilon (\beta-i)\big) = \bar \lambda_{i+1} 
\big(\lambda_{i+1} -\epsilon (\beta-i)\big).
\ee
They imply that
for generic values of $\lambda_1, ..., \lambda_m$ the values $\bar
\lambda_1, ..., \bar \lambda_m$ are fixed up to one parameter. 
A further parameter, the value $k$ of a central element, enters the expression
of $\tilde \lambda_a^{[2]}$ in terms of $\lambda_j$.

 In the case of $so(2m+1)$ we have the $m$th condition
\be \label{u1ms}
 \bar \lambda_m \big(\lambda_m +   \half\big) = 0. 
\ee
It fixes $\lambda_m = - \half$ unless $\bar \lambda_m$ vanishes.

For the case  $\bar \lambda_i = 0,\; i=1,
\ldots,m$ we have the following result. 
The sequence of weights $\lambda_1, ...,\lambda_m$ obeys
the $m-1$ three-factor relations
\be \label{3factors}
(\lambda_i - \lambda_{i+1}) \big( \lambda_i + \lambda_{i+1} -2\epsilon  (\beta - i)\big)
$$ $$
\Big( \half(\lambda^2_i + \lambda^2_{i+1}) - \epsilon (\beta-i+1) (\lambda_i +
\lambda_{i+1}) + \epsilon   \lambda_{i+1} 
+ k - \epsilon  \sum_{j=1}^{i-1} \lambda_j + \half (\beta-i)^2 \Big)
= 0, 
\ee
for $i= 1, ..., m-1$. 
In the case of $so(2m+1)$ we have the $m$th three-factor relation
\be \label{3factors0} 
\lambda_m (\lambda_m + 1 ) \Big ( \half \lambda_m (\lambda_m - 1 )  -   
\sum^m \lambda_j + k +\frac{1}{8} \Big) = 0. 
\ee

The structure of a general sequence of allowed weights $(\lambda_1, ..., \lambda_m)$
can be described similar to the linear case as consisting of segments of
equal values (\ref{lambdamu}). Now the relation of the values $\lambda, \mu$ of two adjacent 
segments can be
determined either by the vanishing of the second bracket, then
$ \mu = - \lambda + 2\e (\beta-i_2) $, 
or by the vanishing of the third bracket. In the latter case the value of
the central element $k$ enters the relation.

In general the solutions of the weight conditions can have three
continuous parameters in the cases $so(2m)$ and $sp(2m)$, but only two
in the case of $so(2m+1)$. 

We present two typical solutions of (\ref{3factors}) encountered in the examples
below. 
The first is given by the weight sequence
\be \label{lamb0}
(\lambda, ..., \lambda, 0 ,.., 0) \ee
with the ratio sequence $f_1(u), ..., f_m(u)$,
$$ 1, ..., 1, \frac{u-\half \epsilon \lambda -\half (\beta-i_0)}
{u+\half \epsilon \lambda -\half (\beta-i_0)}, 1, ...,1.  $$
$i_0$ is the position of the last entry $\lambda$ in (\ref{lamb0}) and of the  
non-trivial entry in the ratio sequence.

The second is given by the weight sequence
\be \label{lambmueps} 
(\lambda, ..., \lambda, \mu, \lambda-\e, ..., \lambda-\e)
\ee
with the ratio sequence
 $$ 1, ..., 1,\frac{u + \frac{\epsilon}{2}(\mu-\lambda) -\half
(\beta-i_0+1)}{u - \frac{\epsilon}{2}(\mu-\lambda) -\half
(\beta-i_0+1)}, 1, ...,1. $$
Here the entry $\mu$ stands at the position $i_0+1$ and the non-trivial ratio
at the position $i_0$.

We present three explicit examples. The first is simply obtained from the 
above linear evaluation examples. The second is well known \cite{Re},
corresponding in the case $so(3)$ to the Quantum angular momentum.
The third is not fully explicit and generalizes the second one. 

\vspace{.5cm}

{\it 1. Product of linear $L$ }

We consider $L$ operators constructed by  the product of
two $L$ operators of the linear evaluation,
$ 
L_1(u) = I_1 u +G_1, \  L_2(u) = u I_2 + G_2 $, 
$$ L_{12}(u) = L_1(u) L_2(u+\delta). $$
The resulting $L_{12}$  depends on $\delta$ and on the parameters in the factors.
In the cases $so(2m)$ and $sp(2m) $ we have $L$ operators depending on one
parameter. If we choose  as $L_1, L_2$ such
expressions the resulting $L_{12}(u)$ depends on up to
three parameters.
The weight function ratios multiply,
$ f_{12, \ i}(u) = f_{1, \ i}(u) f_{2, \ i}(u+\delta) $. 

\vspace{.5cm}

{\it 2. Jordan-Schwinger presentations}

In this example $\bar H$, the $\e$-antisymmetric part of $H$, vanishes.
Then 
\be \label{Hs}
H= \half (G^2 + \beta G + k I). 
\ee
The generators $G$ are built from the canonical pairs
\be \label{ddxs}
 [\dd_a, x_b]_{-\epsilon} = \e_{ab} ,
 \ee
in the following way
\be \label{GJSs}
  G_{ab} = -\epsilon( x_a \dd_b -\epsilon x_b \dd_a).
\ee

We consider representations on polynomial functions of $x_a$ on which $\dd_b$ act as
differentiations. The highest weight conditions are fulfilled by
$$ \psi(x) = (x_{-1})^{2\ell}.$$
  
The resulting weight sequence has the form
$$ (-\e 2\ell, 0, ...,0) $$
for all cases. It compares with (\ref{lamb0}) for $i_0=1$ and 
$\lambda= -2\ell \e$. 
In the case of $sp(2m)$  the parameter $2\ell$ is restricted
to the values $0$ or $1$. In the other cases it is a free parameter.

\vspace{.5cm}

{\it 3. Spinorial type }

The second order evaluation Yangian algebra conditions are fulfilled, if
as in the previous example $ H= \half (G^2 + \beta G + k I)$ 
and the matrix elements of $G$ are constrained to obey
\be \label{Ws}
 W_{ab,cd} = G_{ab} G_{cd} + G_{ac} G_{db} + G_{ad} G_{bc} +
G_{cd} G_{ab} + G_{db} G_{ac} + G_{bc} G_{ad} = 0. \ee

The condition is fulfilled for $G$ constructed as in the previous example.

The weight sequences have the form 
$$ (1, ...,1, \mu, 0, ...,0) $$

In the cases $so(2m)$ and $so(2m+1)$ we have $\mu$ as a free parameter.
In the symplectic case the values of $\mu$ are restricted to $0$ or $1$.

For $so(2m)$ and $so(2m+1)$ the weight sequence and the corresponding 
ratio sequence are a particular case of 
(\ref{lambmueps}) with $\lambda=1$. In the case of $sp(2m)$ we have a particular
case of (\ref{lamb0}) with $\lambda =1$.

\section{Representations of  the highest weight}
\setcounter{equation}{0}

For the study of the representations and in particular of the relations
between their weights the statements about the dependence of the 
 $\epsilon$ symmetric part of $L^{(k)}$ in (\ref{defYan}) is important. We include  a
corresponding proposition.

\begin{prop}
The  $\epsilon$ symmetric part of $L^{(k)}$ (\ref{defYan}) is expressed in terms of products of
lower degree generators and central elements. The  $\epsilon$ antisymmetric part of
$L^{(k)}$ can be considered as the independent generators of the Yangian algebra.
\end{prop}

\noindent
{\bf Proof}

\noindent
Consider  the RLL-relation (\ref{rll}) with the fundamental $R$-matrix
given by (\ref{rzW}) at $u=v-\b$. We  obtain
\be \label{kll}
K_{12}L_1(v-\b)L_2(v)=L_2(v)L_1(v-\b)K_{12}.
\ee
In the index notation this reads
\be\label{kll1}
\e^{a_1a_2}\e_{b_1b_2}L^{b_1}{}_{c_1}(u-\b)L^{b_2}{}_{c_2}(u)=
L^{a_2}{}_{b_2}(u)L^{a_1}{}_{b_1}(u-\b)\e^{b_1b_2}\e_{c_1c_2}.
\ee
After multiplication by $\e_{a_2a_1}$ we obtain
\be\label{kll2}
C_{ab}(u)\equiv L_{ca}(u-\b)L^c{}_b(u)=\frac1n\e_{ab}
L_c{}^d(u)L^c{}_d(u-\b)\equiv \e_{ab}c(u),
\ee

One proves that $C(u)$ contains central elements by showing that it
commutes with $L(v)$,
$$
C(u)L_2(v)=L^t_1(u-\b)L_1(u)L_2(v)=L^t_1(u-\b)R^{-1}_{12}(u-v)L_2(v)
L_1(u)R_{12}(u-v)=
$$
$$
L_2(v)R^{-1}_{12}(u-v)L^t_1(u-\b)L_1(u)R_{12}(u-v)=L_2(v)R^{-1}_{12}
(u-v)C(u)R_{12}(u-v)=L_2(v)C(u),
$$
here in the last step we have used (\ref{c}).

We substitute the expansion (\ref{defYan}) in the spectral parameter
$$ 
C(u)  = 
\sum_{p=0}^{\infty} u^{-p} \sum_{s=0}^p \sum_{i=0}^{p-s} \beta^i \frac{1}
{B(s,i+1)} (L^{(s) \ t} L^{(p-s-i)}). 
$$
We consider the terms at $u^{-p}$. The one with the generators of highest
degree $p$ is proportional to the $\epsilon$ symmetric part of $L^{(p)},
L^{(p)}_{ab} + \epsilon L^{(p)}_{ba}$,  and the remaining terms are 
products of generators of lower degree.

\qed

%%%%%%%%%%%%%%%%%%%%%%%%%%%%%%%%%%%%%%%%%%%%%%%%%%%%%%%%%%%%%%%%%%%%%%%%%%%%
The conditions on the vector of the highest weight $|0\rangle$  
have been formulated above (\ref{hwY}), (\ref{hwY0}). The weight functions
$\lambda_i(-u)$ have been introduced as eigenvalues of $L_{-i, i}(u)$ in
action on  $|0\rangle$  and also the dependent weight functions 
$\lambda_a(-u)$ with $a=-i, 0$ (\ref{hwYadd}). We shall compare also with
the highest weight conditions and the involved weight  functions 
of the $g\ell(n)$ type Yangians (\ref{hwYgl}).

We consider the representation space $V$ of the Yangian built on
the highest weight  vector $|0\rangle $ obeying the  conditions (\ref{hwY}), (\ref{hwY0}). 
Now we 
define the space $V_1$ of vectors obeying a subset of the highest weight conditions
with one of the indices equal to $-1$, 
\be \label{V1}
 V_1 =\Big \{ |\alpha\rangle \in V \quad \big|\quad  L_{-1,j}(u) |\alpha\rangle= 0,  
\qquad j\neq1, \quad  L_{-1,-1}(u) |\alpha\rangle= 0
\Big \}.
 \ee
In the case $so(2m+1)$ the condition 
$$ \quad  L_{-1,0}(u) |\alpha\rangle= 0, $$ is to be added.
In analogy to Lemma 5.13 in \cite{AMR05} we have

\begin{prop}
$V_{1}$ is stable under the action of the subset of generators
$L_{ab}(u), a,b \not = \pm 1$ and the action by these generators forms a
representation of the  Yangian of the corresponding type of the reduced rank $m-1$.
\end{prop}

\noindent
{\bf Proof}

\noindent
For the proof we follow \cite{AMR05}, but notice that due to the difference 
in the highest weight
conditions the index value 1 plays here the role of the index value $m$
there.

(\ref{LuLvosp1}) implies
$$ (u-v) (u-v+\beta ) [L_{-1,a}(u),L_{b,c}(v) ] = 
(u-v+\beta ) (L_{b,a}(v) L_{-1, c} (u) - L_{b,a}(u) L_{-1 c} (v) )+ 
$$ 
$$
+(u-v) ( \epsilon \delta_{b,1} (L_{k, a} (u) L_{-k, c}(v) + \epsilon L_{-k, a} (u) 
L_{k, c}(v)+ L_{0,a}(u) L_{0,c}(v)) 
$$ 
$$
- \epsilon \e_{ac} ( L_{b,k}(v) L_{-1, -k}(u) + \epsilon  L_{b,-k}(v) L_{-1, k}(u)
+  L_{b,0}(v) L_{-1, 0}(u) ) ).
$$
In the case $c \not = \pm 1, b\not = 1$ the first two terms on r.h.s vanish
in action on $|\alpha\rangle \in V_1$ and the last term in action on $|\alpha\rangle$ 
is proportional to 
$$ (L_{b1}(v) L_{-1-1}(u) + \epsilon L_{b,-1}(v) L_{-1,1}(u) ) |\alpha\rangle $$
The action of the first of the latter terms vanishes by the condition on $|\alpha\rangle$.
For the second term  we show that 
$  L_{-1,1}(u) |\alpha\rangle \in V_1$, and therefore it vanishes too.

Indeed, for $a \not = \pm 1$ (\ref{LuLvosp1}) implies
$$  (u-v) (u-v+\beta )[L_{-1,+1}(u) , L_{-1,a}(v) ] |\alpha\rangle= $$ $$
(u-v+\beta)
 (L_{-1,+1}(v) L_{-1,a}(u)-L_{-1,+1}(u)L_{-1,a}(v))\alpha\rangle= 0.
$$
This implies 
$  
L_{-1-1}(v) L_{-1,a}(u) |\alpha\rangle = 0 
$.
If $ a= -1$ we have
$$  (u-v) (u-v+\beta )[L_{-1,+1}(u) , L_{-1,-1}(v) ] |\alpha\rangle= $$ $$
(u-v) 1 (L_{-1+1}(v) L_{-1-1}(u) + \e L_{-1-1}(v) L_{-1+1}(u)) |\alpha\rangle
$$
and this implies
$  
L_{-1-1}(v) L_{-1+1}(u) |\alpha\rangle = 0 
$.
In a similar way we show that
$  L_{1,-1}(u) |\alpha\rangle\in V_1$.

\qed

\subsection{Relations between weight functions from $C(u)$}

Consider  the center generating function (\ref{c}), write the matrix
product using the metric in the form (\ref{metric}) 
$$ 
C_{ab}(u) = \sum_k \Big(L_{-k, a}(u-\beta) L_{+k, b}(u) +\epsilon 
L_{+k, a}(u-\beta)L_{-k, b}(u)\Big)+\eta L_{0, a}(u-\beta)L_{+0, b}(u) ,$$
and    consider the action of both sides on $|0\rangle$
using the highest weight conditions (\ref{hwY}), (\ref{hwY0})
and the RLL relation in commutator form (\ref{LuLvosp1}). 
In the $so(2m+1)$ case the additional term is included with $\eta=1$ and 
the other cases correspond to $\eta=0$.
Relations for the weight functions are obtained with $a=i$ and  $b=-i$. 
Let us  introduce the notations 
\be \label{cLambda}
c_{-k,i}(u,\gamma)|0\rangle=[L_{-k,i}(u-\beta),L_{+k,-i}(u-\gamma)] |0\rangle, 
\qquad \Lambda_i(u,\alpha, \gamma)=\lambda_i(-u+\alpha)\lambda_{-i}(-u+\gamma).\!
\ee
The first argument $u$ will be suppressed.
$$ C_{i,-i}(u) = \sum L_{-k,i}(u-\beta) L_{+k,-i}(u) + \epsilon 
 \sum L_{+k,i}(u-\beta) L_{-k,-i}(u) + \eta  L_{0,i}(u-\beta) L_{0,-i}(u),
$$ 
\be \label{Ci-i0} 
C_{i,-i}(u) |0\rangle= \sum_{k<i} [L_{-k,i}(u-\beta) L_{+k,-i}(u)] |0\rangle+ 
\lambda_i(-u+\beta)\lambda_{-i}(u) |0\rangle \ee
$$
=\Lambda_i(\beta, 0) |0\rangle +\sum_{k<i}c_{-k,i}(0) |0\rangle. 
$$
The special case of $C_{0,0}$ is obtained by substituting
the index value $i \to 0$ but the upper limit of the sum as
$i-1 \to m$, 
$$ 
C_{0,0}(u) |0\rangle= \sum_{k=1}^m c_{-k,0}(0) |0\rangle + \Lambda_0(\beta, 0)|0\rangle. 
$$
The commutator at $u-v+ \beta \to 0 $ is to be calculated. 
We write (\ref{LuLvosp1}) in the form 
$$
(\gamma- \beta) \gamma [L_{a,b}(u-\beta), L_{c,d}(u- \gamma)] = \gamma 
(L_{c,b}(u-\gamma) L_{a,d}(u- \beta )-L_{c,b}(u-\beta), L_{a,d}(u- \gamma))+ 
$$ 
$$
(\gamma - \beta) \{ \e_{ac} \sum L_{k,b}(u-\beta) L_{-k, d}(u-\gamma) + 
\e  L_{-k,b}(u-\beta) L_{+k, d}(u-\gamma) - 
$$ 
$$
\e_{db} \sum L_{c, k}(u-\gamma) L_{a, -k}(u-\beta + 
\e  L_{c, -k}(u-\gamma) L_{a, +k }(u-\beta) \}.
$$
With the notations (\ref{cLambda}) the commutation relation implies
$$ 
\gamma c_{-k_1,k}(\gamma)=\sum_{k_2<k} c_{-k_2,k}+\sum_{k_2<k_1}  c_{-k_1,k_2} 
+\Lambda_k(\beta, \gamma)-\Lambda_{k_1}(\beta, \gamma).
$$
We need also the special case with $ c_{-k_1,0}$. It is obtained by the
substitution of $k$ as the index value $k\to 0$ but as the upper
limit of the sum $ k \to m+1$. The r.h.s. will be abbreviated by
$B_{k_1,k}(\gamma) $.

Expanding in $\gamma$ around $\gamma = 0$ we obtain
\be \label{Bkck} 
B_{k_1,k}(0) = 0, \qquad\quad  c_{-k_1,k}(0)= B^{(1)}_{k_1,k}, \qquad\quad 
 c^{(s-1)}_{-k_1,k}= \frac{1}{s} B^{(s)}_{k_1,k} .
 \ee
The superscript $(s)$ denotes the $s$-fold derivative with respect to 
$\gamma$ at $\gamma=0$.

Let us consider the case of evaluation of order $s$, where the expansion
(\ref{defYan}) is truncated at $s$, i.e. $L^{(p)} = 0, p>s$. In this case we
redefine $L(u)$ multiplying it by $u^s$. 
Then $c^{(p)}_{-k_1, k} = 0$ for $ p \ge s $, thus the last
of the equations (\ref{Bkck}) imply relations for $\Lambda_i$. In the case of quadratic
evaluation we prove the following proposition. 
\begin{prop}
In the quadratic evaluation the weight functions obey $m-1$ relations,
which for the index pair $i, i+1$ have the form 
\be \label{Lambda}
\Lambda_i(u,\beta -i+1, 1) =
\Lambda_{i+1}(u,\beta -i+1, 1), 
\ee
In the case $so(2m+1)$ there is the $m$th relation
\be \label{Lambda0}
\Lambda_m(u, \half, 1)  =
\Lambda_0(u,\half, 1) , 
\ee
where 
$$ 
\Lambda_a (u, \alpha, \gamma)=\lambda_a(-u+\alpha ) \lambda_{-a}(-u+\gamma).  
 $$
\end{prop}

\noindent
{\bf Proof}

\noindent
Consider (\ref{Bkck}) and recall that  $B^{(s)}$ is an expression in terms of 
 $c^{(s)}$ and  $\Lambda^{(s)}$. The latter abbreviates the
$s$-fold $\gamma$ derivative of $\Lambda (u,\beta,\gamma) $ at $\gamma=0$.
$$  
c^{(s-1)}_{-k_1,k}=\frac{1}{s}\Big\{ \Lambda^{(s)}_k-\Lambda^{(s)}_{k_1}
+\sum_{k_2<k} c^{(s)}_{-k_2,k}+\sum_{k_2<k_1}c^{(s)}_{-k_1,k_2}\Big\}=
$$ 
$$ 
\frac{1}{s}\Big\{\Lambda^{(s)}_k-\Lambda^{(s)}_{k_1}\Big\}+\frac{1}{s(s+1)} 
\sum_{k_2<k} \Big\{\sum_{k_3<k}c^{(s+1)}_{-k_3,k} + \sum_{k_3<k_2}  c^{(s+1)}_{-k_1,k_3} 
+\Lambda^{(s+1)}_k - \Lambda^{(s+1)}_{k_2}\Big\} + 
$$ 
$$
\frac{1}{s(s+1)}\sum_{k_2<k_1}\Big\{ \sum_{k_3<k_2}c^{(s+1)}_{-k_3,k_2} 
+\sum_{k_3<k_1}  c^{(s+1)}_{-k_2,k_3} 
+  \Lambda^{(s+1)}_{k_2} - \Lambda^{(s+1)}_{k_1}\Big\}.
$$
This may be continued with higher derivatives and more extended sums.
 In the quadratic evaluation
case the expressions with second derivatives are sufficient, because in this
case $c^{(2)}_{-k_1,k}= 0$.

 The sums simplify  for low values of the indices at
$c^{(s-1)}_{-k_1,k}$, e.g.

\be \label{cs12}
c^{(s-1)}_{-1,2} = \frac{1}{s} \{ \Lambda^{(s)}_2 - \Lambda^{(s)}_{1}\}
+ c^{(s)}_{-1,2},
\ee

We obtain
$$ 
C_{-1,1}(u) |0\rangle= \Lambda_1(\beta,0) |0\rangle=\lambda_1(-u+\beta) \lambda_1(-u)|0\rangle, 
\qquad  C_{-2,2}(u) |0\rangle=\big(\Lambda_2(\beta, 0)+c_{-1,2}(0)\big)|0\rangle. 
$$
In the case of evaluation of order $s$ we have  $c^{(s)}_{-i,k}=0$ and the
last term can be calculated iteratively by (\ref{cs12}).
$$  C_{-2,2}(u) |0\rangle =
\big(\Lambda_2(\beta, 0)+\Lambda^{(1)}_2(\beta, 0)-\Lambda^{(1)}_1(\beta, 0)
+c^{(1)}_{-1,2}(0)\big)|0\rangle =$$ $$
\big(\Lambda_2(\beta 0)+\Lambda^{(1)}_2(\beta, 0)-\Lambda^{(1)}_1(\beta, 0)\big)|0\rangle 
+  
\half \left (  \Lambda^{(2)}_2(\beta , 0) - \Lambda^{(2)}_1(\beta , 0) +  c^{(2)}_{-1,2}(0)
\right )|0\rangle .
$$
(\ref{kll2}) implies $C_{-1,1}(u)|0\rangle=C_{-2,2}(u)|0\rangle$. Therefore
in the case of quadratic evaluation we obtain 
$$
\Lambda_1(\beta, 0) + \Lambda^{(1)}_1(\beta, 0) +  \half
\Lambda^{(2)}_1(\beta, 0) =
\Lambda_2(\beta, 0) + \Lambda^{(1)}_2(\beta, 0) +  \half
\Lambda^{(2)}_2(\beta, 0),  
$$
Thus we  have proven (\ref{Lambda}) for the index pairs $1, 2$.
Relying on  Proposition 2 the analogous relation can be written for the
the index pairs $i, i+1, i=1,2,...m-1$. The value of $\beta$ changes in 
reducing to the lower rank as $ \beta \to \beta - (i-1)$.
In this way we have proven (\ref{Lambda}).

In the case of $so(2m+1)$ the iterative reduction according to Proposition 2
results in the last step in the Yangian of $so(3)$ type. In order to proof
the additional condition (\ref{Lambda0}) it is sufficient to consider the
$so(3)$ case.  

The above sums simplify also in the low rank cases, because the sum indices do not
exceed $m$.
Let us consider the case $so(3)$, where $\beta=\half$. 
We abbreviate $\Lambda_a(u, \half, \gamma)$ by $\Lambda_a(\gamma)$.
$$ 
C_{1,-1} |0\rangle=\lambda_1(-u+\half) \lambda_{-1}(u) |0\rangle= C_{0,0}|
0\rangle=\big(\lambda_1(-u+\half)\lambda_{-1}(u)+c_{-1,0}\big)|0\rangle.
$$
In the finite order evaluation the last term can be calculated by
$$ 
c^{(s-1)}_{-1,0} = \frac{1}{s} (\Lambda_0^{(s)}  - \Lambda_1^{(s)} 
+ c^{(s)}_{-1,0}). 
$$
In the case of quadratic evaluation we have  $c^{(2)}_{-1,0} = 0 $ and
obtain a relation analogous to (\ref{Lambda})
$$
\Lambda_1(\beta, 0) + \Lambda^{(1)}_1(\beta, 0) +  \half
\Lambda^{(2)}_1(\beta, 0) =
\Lambda_0(\beta, 0) + \Lambda^{(1)}_0(\beta, 0) +  \half
\Lambda^{(2)}_0(\beta, 0), 
$$
This result for the particular case $so(3)$ leads to the wanted 
condition (\ref{Lambda0}) in the $so(2m+1)$ case 
by replacing the index values $\pm1$ by $\pm m$ referring to
 Proposition 2. 

\qed 

We add the remark that
instead of referring to Proposition 2 the proof can be completed also by
analyzing (\ref{Ci-i0}), (\ref{Bkck}) for general index values. 
We abbreviate $\Lambda_k(u, \beta, \gamma)$ by $\Lambda_k(\gamma)$.
$$ 
C_{k,-k}|0\rangle=\Lambda_k(0) |0\rangle +\sum_{k_1<k}c_{-k_1,k}(0)|0\rangle=
\Lambda_k(0) |0\rangle +\sum_{k_1<k}\Big(\Lambda_k^{(1)}(0)-\Lambda_{k_1}^{(1)}(0)\Big)
|0\rangle + 
 $$ 
 $$
\half  \sum_{k_1<k} \sum_{k_2<k}\Big((\Lambda_k^{(2)}(0) - 
\Lambda_{k_2}^{(2)}(0)\Big) |0\rangle + \half  \sum_{k_1<k}  
\sum_{k_2<k_1}\Big(\Lambda_{k_2}^{(2)}(0) - \Lambda_{k_1}^{(2)}(0)\Big)|0\rangle.
$$
Here we have used $c^{(2)}_{-k_1,k} = 0$ which holds in the second order
evaluation.
$$ 
C_{k,-k}|0\rangle=\big(\Lambda_k(0)+(k-1) \Lambda^{(1)}_k(0)+ 
\half (k-1)^2 \Lambda^{(2)}_k(0)\big) |0\rangle - 
$$ 
$$
 \sum_{k_1<k}\Lambda_{k_1}^{(1)}(0) |0\rangle- \half (k-1)
\sum_{k_1<k}\Lambda_{k_1}^{(2)}(0) |0\rangle 
$$ 
$$
+ \half  \sum_{k_2<k_1<k}\Lambda_{k_2}^{(2)}(0) |0\rangle
- \half  \sum_{k_1<k}(k_1-1) \Lambda_{k_1}^{(2)}(0) |0\rangle .
$$ 
We use the fact that here the Taylor expansion in $\gamma$ terminates  
here at the second order.
$$  
C_{k,-k}|0\rangle=\Lambda_k(k-1)|0\rangle+\sum_{k_1<k}\big(\Lambda_{k_1}
(0) |0\rangle-\Lambda_{k_1}(1)\big)|0\rangle-\half (k-2)\sum_{k_1<k} 
\Lambda_{k_1}^{(2)}(0) |0\rangle
$$ 
$$
+ \half  (k-2) \Lambda^{(2)}_1 |0\rangle 
- \half (k-2) \Lambda^{(2)}_{k-1} (0)  |0\rangle +
\half \sum_{k_1=2}^{k-2} (k-2k_1) \Lambda^{(2)}_{k_1} (0) |0\rangle =
$$ 
$$
\Lambda_k(k-1)|0\rangle+\sum_{k_1<k}\big(\Lambda_{k_1}(0)-\Lambda_{k_1}
(1)\big)|0\rangle - (k-2) \Lambda^{(2)}_{k-1} (0)|0\rangle- 
 \sum_{k_1=2}^{k-2}\Lambda^{(2)}_{k_1}(0)|0\rangle .
$$
At $k=2$ the last two terms vanish, at $k=3$ the last term vanishes.
$$ 
C_{3,-3}|0\rangle=\big(\Lambda_3(2)+\Lambda_2(0)-\Lambda_2(1)+
\Lambda_1(0)-\Lambda_1(1)-\Lambda^{(2)}_{2} (0)\big)|0\rangle=
 $$ 
 $$
\big(\Lambda_3(2)-\Lambda_2(2)+\Lambda_2(1)-\Lambda_1(1)+ \Lambda_1(0)\big)|0\rangle. 
$$
We obtain that the equations
$C_{1,-1}|0\rangle=C_{2,-2}|0\rangle= C_{3,-3}|0\rangle$ can be written as
$$ 
\Lambda_1(u, \beta, 1)=  \Lambda_2(u, \beta, 1), \qquad  \Lambda_2(u, \beta, 2)=  
\Lambda_3(u, \beta, 2) .
$$
We have restored the suppressed first two arguments. Now we use the fact
that a shift in the spectral parameter $u$ in $L(u)$ results in the
equivalent 
Yangian algebra. With the shift $u \to u-1$ owing to the definition of
$\lambda(u, \alpha, \beta)$ we obtain from the second equation  
$ \Lambda_2(u, \beta-1, 1) = \Lambda_3 (u, \beta-1, 1)$.
This confirms that (\ref{Lambda}) generalizes from the index pairs $1,2$ to
$2,3$ as described above according to Proposition 2. 
This shows another way to prove the proposition.

The relations (\ref{Lambda}),(\ref{Lambda0}) generalize easily
 for higher order evaluations.
The involved $\gamma$ derivatives at $\gamma=0$ constitute the
Taylor expansion with $\gamma=1$. 
In this way we obtain for arbitrary order $s$ evaluations  
the analog of the Proposition 5.14 in \cite{AMR05}.
\begin{prop}
In the general case of highest weight representations of the
orthogonal and symplectic Yangians
the weight functions (\ref{hwY}), (\ref{hwY0}), (\ref{hwYadd})
obey 
\be \label{i,i+1}
 \frac{\lambda_i(u)}{\lambda_{i+1}(u)} = 
 \frac{\lambda_{-i-1}(u-\beta+i)}{\lambda_{-i}(u-\beta +i)}.
\ee
Here $i=1, ..., m-1$ in the cases of $so(2m), sp(2m)$ and $so(2m+1)$.
But in the latter case we obtain the $m$th  relation
\be \label{0,m}
\frac{\lambda_0(u)}{\lambda_{m}(u)} = 
 \frac{\lambda_{-m}(u+\half)}{\lambda_{0}(u+\half)}.
\ee

\end{prop}

Our analysis above was started from the center generating function $C(u)$ and 
has used (\ref{LuLvosp1}) expanded  at $u-v+\beta=0$. It allows to prove this
proposition for arbitrary finite order evaluations. 
Following  \cite{AMR05},
by considering (\ref{LuLvosp1}) at $u-v+\beta=1$ a proof avoiding the Taylor
expansion can be done. The index value 1 plays here the analogous role of the value $m$ in
\cite{AMR05}.

\noindent
{\bf Proof}

\noindent
We start with the commutator $[L_{-1,2}, L_{1,-2}] $ in action on $|0> $.
$$ (u-v) (u-v+\beta) L_{-12}(u) L_{1-2}(v) |0> = 0 + 
(u-v) \{ \e (L_{k,2}(u) L_{-k,-2}(v) + $$ $$ \e L_{-k,2}(u) L_{k,-2}(v))
- \e ( L_{+1,k}(v) L_{-1,-k}(u) + \e  L_{+1,-k}(v) L_{-1,k}(u) ) \} |0> =$$
$$
(u-v) \{  \e L_{-1,2}(u) L_{+1, -2}(v)  +    \e L_{-2,2}(u) L_{+2, -2}(v)  
-  L_{+1,-1}(v) L_{-1,1}(u) \} |0> $$
In the $so(2m+1)$ case the terms in the sums with sum index $0$ do
not give extra contributions.
We obtain
$$ (u-v+\beta -1) L_{-1, 2}(u) L_{1,-2}(v) |0> = 
\lambda_2(u) \lambda_{-2}(v) - \lambda_{-1} (v) \lambda_{-1}(u) 
$$
Choosing $v= u+\beta -1$ we find
$$ \frac{\lambda_1(-u)}{\lambda_2(-u)} = 
 \frac{\lambda_{-2}(-u- \beta+1)}{\lambda_{-1}(-u-\beta+1)}
$$
According to Proposition 2 we consider the subspace $V_1$ obeying (\ref{V1}).
The action of
 $L_{ab}, a,b \not = \pm 1$ forms a representation of the rank $m-1$
Yangian with the same weights besides $\lambda_{\pm 1}(u)$. 
By the same argument the analogous relation involving $\lambda_{\pm 2},
\lambda_{\pm 3}$ is obtained. Proceeding this way  we obtain (\ref{i,i+1}).

To derive (\ref{0,m}) first for the case $ so(3)$ 
 consider the commutator $[L_{-1,0}(u), L_{1,0}(v) ] $. 
The range in the sum over $k$ is merely $k=1$ and $\beta = \half$.
$$  (u-v) (u-v+\beta) [L_{-1,0}(u), L_{1, 0}(v)] |0> = 
(u-v) ( L_{-1,0}(u) L_{1,0}(v) |0> + \lambda_0(u) \lambda_0(v) -
\lambda_{-1}(v) \lambda_{1} (u) )$$
We obtain
$$ \frac{\lambda_0(-u)}{\lambda_{1}(-u)} = 
 \frac{\lambda_{-1}(-u+\half)}{\lambda_{0}(-u+\half)}
$$ 
In the case $so(2m+1)$ this situation appears in the last step of the
reductive iteration according to Proposition 2. This means, for $so(2m+1)$ 
the same results holds with the index value $1$ replaced by $m$.

\qed

\subsection{The simple weight function ratios}

In (\ref{fi}) we  have defined weight function ratios characterizing the 
highest weight representations. We are going to explain how
each of these $m$ ratios can be related to
the weight function ratio of an involved $g\ell(2)$ type Yangian
representation.

We consider the algebra  $\mathcal{Y}^{-i,i+1}$ generated by
$$
L_{-i,i+1}(u),\qquad L_{-i-1,i}(u),\qquad  L_{-i-1,i+1}(u), \qquad L_{-i,i}(u), 
$$
 corresponding term by term to the basic $\mathcal{Y}(g\ell(2))$
generator functions 
\be \label{sl2order}
 L^{g\ell}_{1,2}(u),\qquad L^{g\ell}_{2,1}(u) ,\qquad L^{g\ell}_{2,2}(u),
\qquad L^{g\ell}_{1,1}(u),  
 \ee  
ordered with the one appearing in the highest weight condition (\ref{hwYgl}) 
 first, the excitation operator second and further as they appear in the
commutation relations. Normalization ambiguities are avoided by 
observing that the weight functions of the trivial representation coincide 
being equal to $u$.  Representations of $\mathcal{Y}(g\ell(2))$ on the 
highest weight vector obeying (\ref{hwYgl}) are characterized by the ratio 
of the weight functions
\be \label{fgl2}
 f(u) = \frac{\lambda^{g\ell}_1(u)}{\lambda^{g\ell}_2(u)}. 
 \ee 
In the case of evaluation of order $s$ this function is a ratio of polynomials 
and equal 1 at infinity. 
If such a function  is given then $L^{g\ell}(u)$ and a highest weight
representation can be constructed such that
it is the weight function ratio ( compare \cite{AMR05}, Lemma 5.5).    
Indeed, let $u_k^+, u_k, k=1,..,s$ be the zeros and poles of $f(u)$.
We choose $2s$ Heisenberg canonical pairs,
$$ 
x_{\alpha}^{(k)},\;\; \dd_{\alpha}^{(k)},\quad  k=1,..,s,\quad \alpha = 1,2;  
\qquad\qquad [\dd_{\alpha}^{(k)},  x_{\beta}^{(l)} ] = \delta^{k,l}
\delta_{\alpha,\beta}. 
$$
The  linear $L$ operators obeying the RLL relation (\ref{rll}) with
fundamental $R$ matrix substituted as $R(u) = I u + P$
can be constructed in the Jordan-Schwinger form 
as the $2\times 2$ matrix with the elements
$$ \big(L^{JS}_k(u)\big)_{\alpha,\beta} = u \delta_{\alpha,\beta} - 
x_{\alpha}^{(k)} \dd_{\beta}^{(k)}. $$

Compared to the general expansion of $L(u)$ in inverse powers of $u$ as in 
(\ref{defYan}) we prefer the form multiplied by the order of evaluation,
here by $u$.

The wanted highest weight representation is  spanned by the monomials of 
$x_{\alpha}^{(k)}$ and created on the highest weight vector
$$  |0\rangle = \prod_{k=1}^s (x_1^{(k)})^{u^+_k-u_u} $$
by the generators of the $L$ operator obtained by the matrix product
$$ L^{g\ell}(u) = \prod_{k=1}^s L^{JS}_k(u-u_k). $$

We consider the representation of  $\mathcal{Y}^{-i,i+1}$ on the highest
weight vector $|0\rangle$ of the considered $\mathcal{Y}(\mathcal{G})$ representation. 

We see that $f_i(u),\quad i=1, ...,m-1$,  coincide with the ratio $f(u)$
characterizing the corresponding representation of the Yangian
 $\mathcal{Y}^{-i,i+1}$ of $g\ell(2)$ type.

\be \label{fifgl}
f_i(u) =  f(u),        \quad i=1, ...,m-1 \ee
 
After having considered the ratio $f_1(u)$ in this way we may reduce to the
subspace $V_1$ as described in Proposition 2 and turn to $f_2(u)$ as the
first weight function ratio of the reduced rank $m-1$ representation. 
The reduction proceeds in this way down to $f_{m-1}$. The ratio $f_m(u)$ and
its relation to a $g\ell(2)$ type Yangian representation is to be considered
in more detail.

Considering other contained $g\ell(2)$ type Yangians we find other
weight function ratios. They are related to the ones  
corresponding to the simple roots   by multiplication. 
For example the  $g\ell(2)$ type Yangian representation $\mathcal{Y}^{-1,3}$
corresponds to the ratio
$$  
\frac{\lambda_{1}(u)}{\lambda_{3}(u) } = f_1(u) f_2(u). 
$$

Thus we have associated a $g\ell(2)$ type Yangian representation to each of
the ratios $f_i(u), i=1,\ldots, m-1$. How to associate a 
$g\ell(2)$ type Yangian representation to $f_m(u)$ will be considered in the 
following.

\subsubsection{The even-dimensional orthogonal case}

\begin{prop}
The weight function ratio $f_m(u)$ (\ref{fi}) coincides with the one of the
$g\ell(2)$ type Yangian representation, 
$$ f_m(u) = f(u), $$
generated by the matrix elements of $L(u)$
$$ 
L_{-m,1-m}(u),\qquad L_{m-1,m}(u),\qquad L_{m-1,1-m}(u), \qquad
L_{-m,m}(u), 
$$
with the ordering of the generators as in (\ref{sl2order}). 
\end{prop}

\noindent
{\bf Proof}

\noindent
In the $so(4)$ case we see the following two $g\ell(2)$ Yangian subalgebras.
The first one generated by
$$ 
L_{-1,2}(u),\qquad L_{-2,1}(u),\qquad L_{-2,2}(u),\qquad L_{-1,1}(u), 
$$
is of the type $\mathcal{Y}^{-,+}$ and implies the particular case $i=1$ of
(\ref{fi}).  
The second is generated by
$$ 
L_{-2,-1}(u),\qquad L_{1,2}(u),\qquad L_{1,-1}(u),\qquad L_{-2,2}(u). 
$$
Note that we have observed the ordering of the generators as in
(\ref{sl2order}).
The latter $g\ell(2)$ Yangian subalgebra  leads to the relation
$$ 
\frac{\lambda_{2}(u)}{\lambda_{-1}(u) } = f_2(u) = f(u). 
$$
This is the $m$-th relation (\ref{fi}) in the particular case of rank $m=2$.

We complete the proof of the assertion by referring to Proposition 2. 
 
\qed

\subsubsection{ The symplectic case}

\begin{prop}
$L_{a,b}(2u)$ with $a,b= \pm m$ generate on $|0\rangle$ a 
  $g\ell(2)$ type Yangian representation. Relating the matrix indices
$a,b$  to the indices $\alpha, \beta$ of $g\ell(2)$ as  
$$
a=-m \to \alpha = 1, \qquad a= +m \to \alpha = 2, \qquad  
b=-m \to \beta = 2, \qquad b= +m \to \beta = 1 ,
$$
the weight function ratio $f_m(u)$ (\ref{fi}) is related to that
$g\ell(2)$ weight function ratio $f(u)$ as 
$$ 
f_m(u) = f(\half u). 
$$ 
\end{prop}

\noindent
{\bf Proof}

\noindent 
In the case of $sp(2)$ we show that the commutation relation (\ref{LuLvosp1})
 can be rewritten in the form (\ref{rllgln}).

The index range is here $a,b,= -1, +1$,
 we abbreviate the indices as $+,-$
and write the commutation relations in detail. 
The sum over $k$ in (\ref{LuLvosp1}) consists of one term $ k=1$.
For example:
$$ 
(u-v) (u-v+\beta) [L_{-+}(u), L_{++}(v)] =  (u-v+\beta)
(L_{++}(v) L_{-+}(u) - (L_{++}(u) L_{-+}(v) )) + 
$$ 
$$
(u-v) \epsilon[\epsilon^2L_{++}(u) L_{-+}(v)+\epsilon L_{-+}(u)L_{++}(v)]
$$
 results in 

$$ (u-v) (u-v+\beta-1) L_{-+}(u) L_{++}(v) = 
(u-v+1) (u-v+\beta) L_{++}(v) L_{-+}(u) 
$$ 
$$
- [(u-v)(1-\epsilon) + \beta ] L_{++}(u) L_{-+}(v). 
$$
In the $sp(2) $ case we have 
$\epsilon= -1 , \beta = 2 $, and thus $  [(u-v)(1-\epsilon) + \beta ] = 2 (u-v+1) $.

$$ [L_{++}(u), L_{--}(v)] = \frac{2}{u-v} (L_{-+}(v) L_{+-}(u)- L_{-+}(u) L_{+-}(v)),
$$
$$ [L_{-+}(u), L_{+-}(v)] =  \frac{2}{u-v}
(L_{++}(v) L_{--}(u)- L_{++}(u) L_{--}(v) ), 
$$
$$ L_{-+}(u) L_{++}(v) = \frac{u-v+2}{u-v}  L_{++}(v)L_{-+}(u)
- \frac{2}{u-v}  L_{++}(u)L_{-+}(v),  $$

$$  L_{+-}(v) L_{++}(u) = \frac{u-v+2}{u-v}  L_{++}(u)L_{-+}(v)
- \frac{2}{u-v}  L_{++}(v)L_{-+}(u).  $$
We write also the last two of these relations as commutators 
$$ [L_{-+}(u), L_{++}(v)] =  \frac{2}{u-v}
(L_{++}(v) L_{-+}(u)- L_{++}(u) L_{-+}(v) ), 
$$
$$ [L_{++}(u), L_{+-}(v)] =  \frac{2}{u-v}
(L_{++}(v) L_{+-}(u)- L_{++}(u) L_{+-}(v) ), 
$$
and observe coincidence with the relations for the $g\ell(2)$ case after
substituting the spectral parameter as $u \to 2u $.
$$ L^{sp(2)} (2u) = L^{g\ell(2)} (u). $$
The comparison of the fundamental $R$ matrices reads correspondingly
$$ R^{sp(2)}(u) = R^{g\ell(2)} (\frac{u}{2}), $$
which has been noticed e.g. in  \cite{AMR05} and \cite{KK19}.

Thus  we have the generators
\be \label{L11} 
L_{-1,-1}(u),\qquad L_{+1,+1} (u),\qquad L_{+1,-1}(u),\qquad L_{-1,+1}(u), 
\ee
ordered as in (\ref{sl2order}), and the algebra 
coincides with the one of $g\ell(2)$ for $u \to \half u$. 
In the  trivial representation the weight functions are 
$\lambda_1(u) = \lambda_2(u) = u$ in the $g\ell(2)$ case but
$\lambda_1 (u) = -u, \lambda_{-1} = u $ in the $sp(2)$ case.

This implies
\be \label{lambdasp2} 
\lambda^{sp(2)}_{-1}(u) = \lambda_2^{g\ell(2)}(\half u), \qquad
\lambda^{sp(2)}_{1}(u) = - \lambda_1^{g\ell(2)}(\half u).
\ee
In the case of $sp(2m)$ in the last step of the iterative reduction in the
sense of Proposition 2
we have the $sp(2)$ type Yangian algebra with generators like (\ref{L11}) but
with the indices $\pm 1$ replaced by $\pm m$. Therefore we have to substitute
also in (\ref{lambdasp2}) $\lambda^{sp(2)}_{\pm1}(u)$ by
$\lambda^{sp(2m)}_{\pm m}(u) $ and thus
$$ f_m(u) = - \frac{\lambda^{sp(2m)}_{m}(u)}{\lambda^{sp(2m)}_{-m}(u)} =
\frac{ \lambda_1^{g\ell(2)}(\half u)}{ \lambda_2^{g\ell(2)}(\half u) } 
 = f(\half u). $$
\qed

\subsubsection{The odd-dimensional orthogonal case}

\begin{prop}
A $g\ell(2)$ type Yangian representation generated by 
$L^{g\ell}(2u)$ results in a representation of the $so(3)$ type Yangian
generated by $L_{ab}(u), a,b= \pm 1, 0$ obtained from the fusion relation
\be \label{LsigmaL}
 L_{a b}(u) = \half  (\sigma_{a})_{\d}^{\a}\big(L^{g\ell}\big)^ \a_{\ \;\b}(2u)
 (\sigma_{b})^{\b}_{\c}\big((L^{g\ell})^{-1}\big)^{\c}_{\;\;\d}(2u+2).
\ee
Any weight function ratio $f_m(u)$ (\ref{fi}) of the  
$so(3)$ type Yangian representation generated by $L_{ab}(u),\;\; a,b= \pm 
m, 0$  in action on the same highest weight vector $|0\rangle$ can 
obtained from a  $g\ell(2)$ type weight function ratio as  $$ f_m(u) = f(2u). $$
\end{prop}

\noindent
{\bf Proof}

\noindent
A general $L$ operator of the orthogonal type Yangians can be obtained from the
 spinorial monodromy $T$ by fusion, where 
$T(u)$ obeys the analog of the
$RLL$ relation (\ref{rll}) with the fundamental $R$ matrix replaced 
by the spinorial one
$\mathcal{R}(u)$ \cite{KK19} and $L$ replaced by $T$.
 The indices $a_1b_1,a_2b_2$ in  (\ref{rll}) labelling  a basis of the
fundamental representation $V_f$ have to be replaced by 
$\alpha_1, \beta_1, \alpha_2, \beta_2$ labeling a basis of the spinor
representation. In the case $so(3)$ the fundamental representation is 3
dimensional and the spinor representation 2 dimensional. In this case
the spinorial $\mathcal{R}$ matrix is simply related to the fundamental $R$ matrix of
$g\ell(2)$  \cite{Re91} 
$$ 
\mathcal{R}^{so(3)}(u) = R^{g\ell(2)}(2u) = 2u I + P,
$$
here the $P$ denotes the permutation operator, its matrix elements are as in
(\ref{IKP}). 

The relation between the $R$ matrices implies for the monodromies a similar
relation,
$$ T(u)  = L^{g\ell(2)}(2u). $$

Further, we have the fusion relation between the spinorial
$T^{\alpha}_{\beta} $ and fundamental $L^a_b $ monodromies, 

\be\label{t2t1}
T^{\a}_{\;\b}(u+\b-\frac12)(\c_{b})
^{\b}_{\c}(T^{-1})^{\c}_{\;\;\d}(u+\b+\frac12)
=L^{a}_{\;\;b}(u)(\c_{a})_{\d}^{\a},
\ee
where in the $so(3)$ case $\beta = \half$,  the vector indices run over 3 values, $a,b= -1,0
+1$ and the spinor indices over 2 values, $\alpha, \beta = 1,2$.
The gamma matrices are just the Pauli matrices. In the standard form
$\sigma_{\pm}$ have just one non-vanishing component,  and $ \sigma_0 =diag(1,-1)$ has two.

 We obtain the fusion relation between the $g\ell(2)$ monodromies
$L^{g\ell}(u)$
and the fundamental $so(3)$ monodromies $L(u)$ (\ref{LsigmaL}).
We have e.g.
$$ 
L_{-1,-1}(u)=\big(L^{g\ell}\big)^{1}{}_2(2u)\big((L^{g\ell}\big)^{-1})^{1}{}
_2(2u+2),
$$
$$
L_{0,-1}(u) =\big(L^{g\ell}\big)^{1}{}_2(2u)\big((L^{g\ell})^{-1}\big)^{1}{}
_1(2u+2)-
\big(L^{g\ell}\big)^2{}_2(2u) \big((L^{g\ell})^{-1}\big)^{1}{}_2(2u+2), 
$$
$$ 
L_{+1,-1}(u) =\big(L^{g\ell}\big)^{2}{}_2(2u)\big((L^{g\ell})^{-1}\big)^{1}{}_1(2u+2), 
$$
$$ 
L_{-1,+1}(u) =\big(L^{g\ell}\big)^1{}_1(2u) \big((L^{g\ell})^{-1}\big)^{2}{}_2(2u+2), 
$$ 
$$ 
L_{0,0}(u) = \half \left ( 
 \big(L^{g\ell}\big)^1{}_1(2u) \big((L^{g\ell}\big)^{-1})^{1}{}_1(2u+2) 
+\big(L^{g\ell}\big)^ 2{}_2(2u)\big((L^{g\ell})^{-1}\big)^2{}_2(2u+2)\right.   
$$ 
$$ 
\left. -\big(L^{g\ell}\big)^2{}_1(2u) \big((L^{g\ell}\big)^{-1})^1{}_2(2u+2)
-\big(L^{g\ell}\big)^1{}_2(2u)\big((L^{g\ell})^{-1}\big)^2{}_1(2u+2)\right ). 
$$
$(L^{g\ell})^{-1}$, the inverse matrix of $(L^{g\ell })$,  is calculated by
$$ (L^{g\ell})^{-1} = \big(qdet L^{g\ell}(u-1)\big)^{-1} 
\begin{pmatrix} 
L^{g\ell}_{22}(u-1) & - L^{g\ell}_{12}(u-1) \\
- L^{g\ell}_{21}(u-1) & L^{g\ell}_{11}(u-1) 
\end{pmatrix}, 
$$
$$ qdet L^{g\ell} (u-1) = L^{g\ell}_{11}(u) L^{g\ell}_{22}(u-1) - L^{g\ell}_{21}(u) 
L^{g\ell}_{12}(u-1). $$
We check that the highest weight conditions for $g\ell(2)$ (\ref{hwYgl}) imply the
highest weight conditions (\ref{hwY}), (\ref{hwY0}) for the resulting $L(u)$. 
In particular we obtain the weights
$$ L_{-1,+1}(u) |0\rangle= \lambda_1(-u) |0\rangle= qdet^{-1} \ \lambda^{g\ell}_1(-2u) 
\lambda^{g\ell}_1(-2u-1) |0\rangle,
$$
$$ \lambda_{-1}(-u) = qdet^{-1}  \  \lambda^{g\ell}_2(-2u)
\lambda^{g\ell}_2(-2u-2), $$
$$ 
\lambda_{0}(-u) = qdet^{-1} \   \lambda^{g\ell}_2(-2u)
\lambda^{g\ell}_1(-2u-1). 
$$
$qdet$  stands for the value of the quantum determinant.
In order to obtain the result for $\lambda_0$
the commutator $[L^{g\ell 1}_{\ \ \ 2}(2u), 
L^{g\ell 2}_{\ \ \ 1}(2u+1)  ]$ is calculated according to 
(\ref{rllgln}).
In the weight function ratios the value of the quantum determinant cancels.

In this way we obtain the weight function ratio of the $so(3)$ type Yangian
representation in relation to the $g\ell(2)$ type Yangian representation ratio
$$ f_1(u) = \frac{\lambda_{1}(u)}{\lambda_0(u)} =
\frac{\lambda^{g\ell}_1(2u)}{\lambda^{g\ell}_2(2u)} = f(2u). $$

The reductive iteration according to Proposition 2 leads to a $so(3)$ type Yangian
in the last step with the generators $L_{a,b}, a,b= \pm m, 0 $.
We obtain the wanted relation between the weight function ratios, 
\be \label{m0}
f_m(u) = 
 \frac{\lambda_{m}(u)}{\lambda_0(u)} = 
\frac{\lambda^{g\ell}_1(2u)}{\lambda^{g\ell}_2(2u)} = f(2u).
 \ee

We have shown above that
for any representation with the ratio $f_m(u)$ the corresponding
$g\ell(2)$ type Yangian representation can be constructed. 

\qed

\subsection{Finite-dimensional representations}

We rely on the {\bf Criterion } \cite{VOT84,Drin88} :

{\it An irreducible representation  of highest weight  of the Yangian $\mathcal{Y}(g\ell(2))$
is finite-dimensional if and only if  weight function ratio can be written
as
\be \label{Pgl2}
 f(u) = \frac{\lambda^{g\ell}_1(u)}{\lambda^{g\ell}_2(u)} =
\frac{P(u+1)}{P(u)}. \ee
$P(u)$ is a polynomial in $u$ with the coefficient at the leading  term 
equal to $1$. It characterizes this representation uniquely.
}

We  obtain the  finiteness conditions for the considered Yangian
representations of orthogonal and symplectic types
in terms of the weight function ratios (\ref{fi})
\be \label{Pi} 
 f_i(u) = \frac{\lambda_i(u)}{\lambda_{i+1}(u)} =
\frac{P_i(u+1)}{P_i(u)},\qquad\qquad i=1, ..., m-1 , \qquad\qquad  
 f_m (u) = \frac{P_m(u+\Delta)}{P_m(u)}, \ee
where the definition of the latter ratio (\ref{fi}) and the shift 
$\Delta$ are specific for the cases:
$\Delta=1$ for $so(2m)$, $\Delta= \half $ for $so(2m+1)$ and
$\Delta=2 $ for $sp(2m)$.

We have shown that the $m$ conditions (\ref{Pi}) are necessary 
for a representation module to
contain a finite-dimensional sub-module.  We have obtained the set
of conditions in the form corresponding to our formulation of the highest
weight relations, modifying in details the form obtained in \cite{AMR05}. 
There it is proven that the set of conditions is also sufficient 
for finding a
finite-dimensional representation. This proof consists basically in
constructing finite-dimensional representations for all the  weight
function sets obeying the conditions.

The finiteness conditions can be formulated in terms of the 
positions of zeros and poles of the ratio functions. In the case of
evaluation of order $s$
finite-dimensional representations are encountered if these positions 
obey the conditions that, after appropriate ordering, the differences
$ u^+_{i,k}- u_{i,k} $ are all real,  and
 for $i=1,..., m-1$ non-negative integer and for $i=m$,
specific to the cases, non-negative integer for 
$so(2m)$, non-negative
half-integer for $so(2m+1)$, non-negative
even-integer for $sp(2m)$. 
Then each of the ratios can be written in terms of a 
polynomial as above (\ref{Pi}).

\section{Representations of the quadratic evaluation}
\setcounter{equation}{0}

In the case of quadratic  evaluation, where  the expasion in (\ref{defYan})
 is truncated after the second non-trivial term, $L$ can be written as
(\ref{LGH}).

\subsection{Algebra relations}

The algebra relations are obtained from (\ref{rll}) or from (\ref{LuLvosp1}).
The commutation relations  of the matrix elements of $G$ 
are written above in (\ref{Liealg0})  and the commutation relations between
the matrix elements of $G$ and $H$ are as in (\ref{Lieadj}).
The commutation relation of the matrix elements of $H$ can be derived
from (\ref{LuLvosp1}) in the limit $u \to 0, v\to 0$.
\be \label{HHcomm}
[H_{a_1b_1}, H_{a_2,b_2} ]= H_{a_2b_1} G_{a_1 b_2} - G_{a_2 b_1} H_{a_1b_2}
+ \beta^{-1} \left (\epsilon \e_{a_1a_2} (H^tH) _{b_1b_2} - \e_{b_1b_2} (H H^t)_{a_2a_1} 
\right ).
\ee
 
Further  constraints resulting from the truncation are contained in the
generating function of the center (\ref{c}).
$$
C_{ab}(u)=L_{da}(u-\b)\e^{ed}L_{eb}(u)=\Big((u-\b)^2\e_{da}+(u-\b)G_{da}
+H_{da}\Big)\e^{ed}\Big(u^2\e_{eb}+uG_{eb}+H_{eb}\Big)=
$$
\be\lb{cec}
=u^2(u-\b)^2\e_{ab}+u^2(u-\b)\Big(G_{ab}+\epsilon G_{ba}\Big)+u(u-\b)
\Big(H_{ab}+\epsilon H_{ba}+G_{da}\e^{ed}G_{eb}-\b G_{ab}\Big)+
\ee
$$
+u\Big(H_{da}\e^{ed}G_{eb}+G_{da}\e^{ed}H_{eb}-\b(H_{ab}-\epsilon 
H_{ba})\Big)+\Big(H_{da}\e^{ed}H_{eb}-\b G_{da}\e^{ed}H_{eb}
+\b^2H_{ab}\Big)=
$$
$$
=u^2(u-\b)^2\e_{ab}+u^2(u-\b){\mathbf{C}}^{(2.1)}_{ab}+u(u-\b){\mathbf{C}}_{ab}
^{(2.3)}+u{\mathbf{C}}^{(2.6)}_{ab}+{\mathbf{C}}^{(2.8)}_{ab}.
$$
We recognize in ${\mathbf{C}}^{(2.1)}_{ab}={\bf{c}}^{(2.1)}\e_{ab},\;\;$ 
${\mathbf{C}}^{(2.3)}_{ab}={\bf{c}}^{(2.3)}\e_{ab},\;\;$ ${\mathbf{C}}^{(2.6)}
_{ab}={\bf{c}}^{(2.6)}\e_{ab}\;\;$ and ${\mathbf{C}}^{(2.8)}_{ab}={\bf{c}}^{(2.8)}\e_{ab}\;\;$
the symmetric constraints deduced in \cite{KK17}  from the RLL Yang-Baxter 
relation and labelled there as the constraints no. 1, 3, 6, 8.

$$ G + \epsilon  G^t = c^{(2,1)} I ,  $$
$$ H + \epsilon  H^t +   \epsilon  G^t G - \beta G =  c^{(2,3)} I, $$
$$ \epsilon  H^t G + \epsilon  G^t H  - \beta (H-\epsilon  H^t) =   c^{(2,6)}
I,  $$
\be \label{symC1368}
 \epsilon  (H- \beta G)^t H + \beta^2 H = c^{(2,8)} I. \ee
A shift in the argument of $L(u)$ results in the equivalent algebra and can
be applied to obtain  a vanishing $c^{(2,1)}$. Then the first condition in
(\ref{symC1368}) reduces to the statement that the matrix $G$ is
$\e$- antisymmetric.

The relations of the case of the linear evaluation are obtained from the 
above by the substitution $H=0$. The first relation remains unchanged. The
second relation results in the second order polynomial condition  on the
matrix $G$,
\be \label{G2betaG}
 G^2 + \beta G = c_2 I, \ \ n c_2 = tr G^2. \ee

\subsection{The conditions on the representation weights}

We have considered relations of the weight functions following
from the generating function of the center $C(u)$. 
They are appropriate to formulate compactly the restrictions on the weight
functions implied by truncation of (\ref{defYan}). 

We have 
(\ref{Lambda}) and the analogous relations for the index values
$1,2$ replaced by $i, i+1$, $i= 1,\ldots,m-1$ as formulated in Proposition 3.

We write the weight functions decomposed in powers of $u$ in analogy to
(\ref{LGH}),  $\lambda_i^{[1]}$ and $\lambda_i^{[2]}$
are the eigenvalues of $G_{-i,i}$ and $H_{-i,i}$ in action on
the highest weight  vector $|0 \rangle$. Thus weight functions have the form
(\ref{lambu2}) and (\ref{Lambi}) reads
$$ \Lambda_{i}(u,\gamma) = \left ((u-\beta)^2 \e + (u-\beta) \lambda_i^{[1]} +
\lambda_i^{[2]} \right) \left ( (u-\gamma)^2 + (u-\gamma)  \lambda_{-i}^{[1]} + 
\lambda_{-i}^{[2]} \right).
$$
We  substitute to (\ref{Lambda}),  
\be \label{lambdaii+1}
 \left ((u-\beta+i-1)^2 \epsilon  + (u-\beta+i-1) \lambda_i^{[1]} +
\lambda_i^{[2]} \right)
\left ( (u-1)^2 + (u-1)  \lambda_{-i}^{[1]} + \lambda_{-i}^{[2]} \right)
= \ee $$
\left ((u-\beta+i-1)^2 \epsilon  + (u-\beta+i-1) \lambda_{i+1}^{[1]} +
\lambda_{i+1}^{[2]} \right)
\left ( (u-1)^2 + (u-1)  \lambda_{-i-1}^{[1]} + \lambda_{-i-1}^{[2]}
\right).
$$
We have $m-1$ relations of this type restricting the
components of the weight functions.  
In the case $so(2m+1)$ we have the additional $m$th relation
\be \label{lambdam0}
 \left ((u-\half)^2 \epsilon  + (u-\half) \lambda_m^{[1]} +
\lambda_m^{[2]} \right)
\left ( (u-1)^2 + (u-1)  \lambda_{-m}^{[1]} + \lambda_{-m}^{[2]} \right)
= \ee 
$$
\left ((u-\half)^2   +
\tilde \lambda_{0}^{[2]} \right)
\left ( (u-1)^2   + \tilde \lambda_{0}^{[2]} \right).
$$
We have used $ \lambda_0^{[1]}= 0, \ \bar \lambda_0^{[2]}= 0$.

\subsubsection{The linear evaluation}

\begin{prop}
The representations of the linear evaluations of the Yangian algebra
are characterized by the weights $\lambda_i$, $i=1, ..,m$ obeying
for $i=1, ..,m-1$
\be \label{lambdalin}
(\lambda_{i+1} - \lambda_i) \big(\lambda_{i+1} +
\lambda_i- \epsilon (\beta-i)\big) = 0. 
\ee
and in the case $so(2m+1)$ the additional $m$th relation
\be \label{lambdalin0}
 \lambda_m \big( \lambda_m + \half\big) = 0 .
\ee

\end{prop}

\noindent
{\bf Proof}

\noindent
In the linear evaluation case we have to substitute in (\ref{lambdaii+1})
and (\ref{lambdam0})   $ \lambda_a^{[2]}= 0$. We abbreviate $ \lambda_i^{[1]}=
\lambda_i$. A factor $(u-1)(u-\beta+i-1)$ can be cancelled.
We expand in powers of $u$ and observe that only the terms independent of
$u$ result in conditions on $\lambda_i$, which can be written in the above
form.
\qed

In the case $so(2m+1)$ 
we obtain two solutions  for  $\lambda_m $. Choosing one of them
the relation (\ref{lambdalin}) at $i=m-1$ has two solutions for 
 $\lambda_{m-1}^{[1]} $. Continuing we get a set of $2^m$ 
$m$-tuples of allowed weights. 

For example, the sequence of allowed weights
$$ (\lambda_1, ...,\lambda_m) = (-\half, ..., -\half) $$
corresponds to the sequence of weight function ratios
\be \label{-half0}
 f_1(u), ..., f_{m-1}, f_m(u)  = 1, ..., 1, \frac{u+\half}{u}  \ee
and the further example of a weight sequence solving (\ref{lambdalin}) and 
(\ref{lambdalin0}),
$$ (-\half, ...-\half,0), $$
corresponds to
$$ 1,...,1, \frac{u+\half}{u}, 1. $$ 
Thus in the case $so(2m+1)$ 
there is no continuous parameter in the allowed sequences of weights.

This is different in the other cases, the linear evaluations of 
 $so(2m)$ or $sp(2m)$ type Yangians, where the allowed representations have 
one continuous parameter  in their weights. 
In particular allowed weights may be 
$$(\lambda_1, \lambda_2, \ldots,  \lambda_m) = 
(\lambda,   \ldots, \lambda,  \lambda), $$
where $\lambda$ can take any complex value.
The corresponding sequence of weight function ratios is
\be \label{lambdadots}
1, \ldots, 1,  \frac{u-\epsilon \lambda}{u+\epsilon \lambda}. 
\ee
Let us consider another example of a weight sequence solving (\ref{lambdalin}),
in the  $so(2m)$ case
\be \label{lambdadots-}
 (\lambda, \ldots, \lambda, -\lambda) \ee
and the corresponding weight function ratios,
$$  1, \ldots, 1, \frac{u- \lambda}{u+ \lambda}, 1 . $$
In the $sp(2m)$ case we have for example the allowed weight sequence
\be \label{lambdadots-2}
 (\lambda, \ldots, \lambda, -\lambda -2) \ee
 and the corresponding weight function ratios,
$$   1, \ldots, 1, \frac{u+ \lambda}{u- \lambda-2},
\frac{u-\lambda-2}{u+\lambda+2} . $$

\subsubsection{The quadratic evaluation}

We analyse (\ref{lambdaii+1}). 
First we observe the particular solution $\lambda^{[1]}_i = \lambda^{[1]}_{i+1}=
\lambda$, $\bar \lambda^{[2]}_i = \bar \lambda^{[2]}_{i+1}= \bar \lambda $.
In the cases $so(2m), sp(2m)$ we have just the  $m-1$ conditions 
(\ref{lambdaii+1}), and this implies the allowed sequences of
weights $\lambda^{[1]}_i$ and $\bar \lambda^{[2]}_i $ of the form
\be \label{lambdabarlambda}
 (\lambda, ..., \lambda); \ \ (\bar \lambda, ..., \bar \lambda) \ee
In the case $so(2m+1)$ we have the $m$th condition (\ref{lambdam0})
restricting the values of the parameters $\lambda, \bar \lambda $.
A particular solution is $\lambda= 0, \bar \lambda=0$ involving the
trivial representation. Other allowed sequences of weights  
are obtained from the following analysis. 

We separate even and odd parts in $ \lambda_{i}^{[2]}$,
$  
\lambda_{i}^{[2]} =  \tilde \lambda_{i}^{[2]} + \bar \lambda_{i}^{[2]}, \qquad
\tilde \lambda_{-i}^{[2]} = \epsilon  \tilde \lambda_{i}^{[2]},\qquad
 \bar \lambda_{-i}^{[2]} =- \epsilon \bar \lambda_{i}^{[2]}. 
 $ 
Let us abbreviate the notation as 
%\be \label{i-i}
$ \lambda_{i}^{[1]} = -\epsilon  \lambda_{-i}^{[1]} 
\to \lambda_i, \qquad 
\bar \lambda_{i}^{[2]} = -\epsilon  \bar \lambda_{-i}^{[2]} \to \bar
\lambda_{i} $. 
We calculate $\tilde \lambda_i^{[2]} $. The second of the constraints
(\ref{symC1368}) implies that it is the eigenvalue of
$\half (G^2 + \beta G + k)_{-i, i} $ in action on $|0\rangle$. Here $k= c^{(2,3)}$. 
Using the
commutation relations (\ref{Liealg0}) and the conditions of highest weight 
(\ref{hwY}), (\ref{hwY0}) we obtain
\be \label{lambda2i} 
 2 \tilde \lambda_{-i}^{[2]} = 2 \epsilon  \tilde \lambda_i^{[2]} =
  \lambda_i (\lambda_i - \epsilon (\beta -i+1))  - \epsilon  \sum_{j<i} \lambda_j + k. 
\ee
In the case $i=1$ we substitute in  (\ref{lambdaii+1})
$$ 2 \tilde \lambda_{-1}^{[2]}  =  \lambda_1^{2} - \beta \epsilon  \lambda_1+
k, \qquad    
2 \tilde \lambda_{-2}^{[2]}  = \lambda_2^{ 2} - \beta \epsilon 
\lambda_2 + k +  \e ( \lambda_{2} - \lambda_{1}). $$  

We decompose (\ref{lambdaii+1}) for $i=1$ in powers of $u$ and
 find that the partial conditions at $u^4$, $u^3$ and $u^2$ are fulfilled
and do not imply restrictions on $\lambda_2, \lambda_1, \bar \lambda_{2},
\bar \lambda_{1} $. This is expected because by the substitutions
(\ref{lambu2}), (\ref{lambda2i})
we have introduced the solutions of the first two of the symmetric
constraints (\ref{symC1368}).
At $u^1$ we have
$$ \bar \lambda_1 \big(\lambda_1 - \epsilon(\beta-1)\big) = \bar \lambda_2 
\big(\lambda_2 -\epsilon(\beta-1)\big).
$$
A particular solution is 
$  \lambda_1 = \lambda_2,  \bar \lambda_1 = \bar \lambda_2 $.
We see that $\bar \lambda_1 = \bar \lambda_2$ implies 
$\lambda_1 = \lambda_2$ unless  $\bar \lambda_1 = \bar \lambda_2= 0$.

For generic values of $i$ (\ref{lambdaii+1}) implies 
the $m-1$ conditions ($i=1, ...,m-1$)
\be \label{u1i}
\bar \lambda_i \big(\lambda_i - \epsilon (\beta-i)\big) = \bar \lambda_{i+1} 
\big(\lambda_{i+1} -\epsilon (\beta-i)\big).
\ee
 In the case of $so(2m+1)$ we have the $m$th condition
$$ \bar \lambda_m \big(\lambda_m +   \half\big) = 0. $$
It fixes $\lambda_m = - \half$ unless $\bar \lambda_m$ vanishes.

Let us analyse in detail the particular case  $\bar \lambda_i = 0,\; i=1,
\ldots,m$. 

\begin{prop}
In the case where in the weight functions of quadratic evaluation
representations all the odd parts in the last term $\lambda_i^{[2]} =
\tilde \lambda_i^{[2]} + \bar  \lambda_i^{[2]} $
vanish, i.e. $\bar  \lambda_i^{[2]} = 0$, the conditions on the weights
$\lambda_i^{[1]} = \lambda_i$ have the factorized form
\be \label{3 factors}
(\lambda_i - \lambda_{i+1}) \big( \lambda_i + \lambda_{i+1} -2\epsilon  (\beta - i)\big)
$$ $$
\Big( \half(\lambda^2_i + \lambda^2_{i+1}) - \epsilon (\beta-i+1) (\lambda_i +
\lambda_{i+1}) + \epsilon   \lambda_{i+1} 
+ k - \epsilon  \sum_{j=1}^{i-1} \lambda_j + \half (\beta-i)^2 \Big)
= 0, 
\ee
for $i= 1, ..., m-1$. 
In the case of $so(2m+1)$ we have the $m$th relation
$$ 
\lambda_m (\lambda_m + 1 ) \Big ( \half \lambda_m (\lambda_m - 1 )  -   
\sum^m \lambda_j + k +\frac{1}{8} \Big) = 0. 
$$
\end{prop}

\noindent
{\bf Proof}

\noindent
The relevant contributions in 
 (\ref{lambdaii+1}) are the ones independent of $u$. 
Consider this relation for $i=1$.
We cancel $(\lambda_1 -
\lambda_2) $ and obtain
$$
\Big( k  + \half (1-\beta)^2 \Big)
  \Big(\frac{\epsilon }{2} (\lambda_1 + \lambda_2) +1 - \beta\Big) =
\frac{\epsilon }{4} ( (\lambda_1 - \lambda_2)- 
\Big( \frac{1}{2}  (\lambda_1^2 + \lambda_2^2) +
$$  
$$
(\lambda_1 + \lambda_2) \epsilon (\half - \beta) \Big) \Big( \frac{\epsilon }{2} 
(\lambda_1+\lambda_2)+(\half-\beta)\Big) -\Big( \frac{1}{2}+\lambda_2^2+
\lambda_2 \epsilon (\half - \beta) \Big) \Big( \frac{\epsilon }{2}  \lambda_2 +
(\half - \beta)  \Big).
$$
Now one can show that the r.h.s. can be divided by the second factor
of the l.h.s. This is expected because $k$ is the value of one of the
central elements. In this way we obtain the factorization for the case
$i=1$. 
We use Proposition 2 and (\ref{lambda2i}) to obtain the result for generic $i$. 
Note that it takes to substitute
$
\beta \to \beta -i+1, \qquad k \to k- \sum_{j<i} \lambda_j $.
 (\ref{lambdam0}) implies the particular form of the $m$th relation in the
case $so(2m+1)$. 

\qed

%%%%%%%%%%%%%%%%%%%%%%%%%%%%%%%%%%%%%%%%%%%%%%%%%%%%%%%%%%
Let us discuss typical sequences of weights $(\lambda_1, ...,\lambda_m)$
solving (\ref{3 factors}) in the cases $so(2m)$ and $sp(2m)$.

The condition (\ref{3 factors}) is fulfilled in a simple way if
the first factor vanishes for all $i$. This results in the allowed sequence
of weights (\ref{lambdabarlambda}) with $\bar \lambda=0$.
It corresponds to the sequence of weight function ratios
\be \label{lambdaldots2}
 1, \ldots, 1, \frac{u^2 - u \epsilon \lambda + \half (\lambda^2 - \beta
\epsilon \lambda + k)}{u^2 + u \epsilon \lambda + \half (\lambda^2 - \beta
\epsilon \lambda + k)} . \ee
Both $\lambda$ and $k$ can take arbitrary values.
We notice that at the special value of $k$, $k= - \lambda^2 +\epsilon \beta
\lambda $, the weight functions (\ref{lambdaldots2}) coincide with the ones
of the corresponding case of linear evaluation (\ref{lambdadots}). 

The vanishing of the second factor at some $i_0$ changes the values for the
next $\lambda_{i_0+1}$ and the following weights. For example the sequence
\be \label{lambdamu}
 (\lambda, ...,\lambda, \mu, ..., \mu), \ee
where the first $i_0$ weights have the equal value $\lambda$ and the
remaining $m-i_0$ weights have all the value $\mu =- \lambda +2 \epsilon
(\beta-i_0) $ is a solution. The change in the sequence is determined by $\beta$,
i.e. is fixed by the rank. The corresponding sequence of weight function ratios
is for $m> i_0+1$
\be \label{lambdamu2}
 1, \ldots, 1, \frac{u^2 - u \epsilon \lambda + \half (\lambda^2 - \beta
\epsilon \lambda + k)}{u^2 - u \epsilon \mu + \half (\mu^2 - \beta
\epsilon \mu + \epsilon i_0 (\mu- \lambda) + k)}, 1, \ldots, 1.   
\ee
The non-trivial ratio stands at the position $i_0$. 

If in the case $so(2m+1)$ the conditions (\ref{3factors}) and
(\ref{3factors0}) are fulfilled by the vanishing of the first two factors
we obtain weight sequences with no free parameter, but  the central
element value $k$ is free.
For example if we choose $\lambda_m=0$ as solution of (\ref{3factors0})
and the vanishing of the second bracket appears at $i_0$ only then we obtain 
the solution of the form (\ref{lambdamu}) with $\mu=0$ and $\lambda= 2
(\beta- i_0)$.  The weight function ratios have the form (\ref{lambdamu2})
with these values and $\e=+1$.

By the vanishing of the third factor in (\ref{3 factors}) 
the amount of change in the weight sequence 
is also determined by the value of the central element $k= c^{(2,3)}$. 
In this way the sequence
\be \label{lambda00}
 (\lambda, 0, ..., 0) \ee
solves the condition with the value of $k$ fixed at
$$ k= - \half (\beta-1)^2 - \half \lambda^2 + \epsilon \beta \lambda. $$
The corresponding weight function ratios $f_1(u), f_2)u), \ldots, f_m(u) $ are
\be \label{flambda00}
 \frac{u-\half \epsilon \lambda -\half (\beta-1)}
{u+\half \epsilon \lambda -\half (\beta-1)}, 1, \ldots, 1, 
\ee
if the rank $m>2$. It holds also for $so(3)$ and $sp(4)$.
In the case $so(4)$ we have $f_1(u)= f_2(u)$ with the above expression for
$f_1(u)$. In the case $sp(2)$ we have
$$f_1(u) = \frac{(2u + \lambda-1)(2u+ \lambda+1)}{(2u - \lambda-1)(2u-
\lambda+1)}.
$$
The finiteness conditions are fulfilled if $-\epsilon \lambda $ is a
non-negative integer,
besides of the case $sp(2)$ where the values of $\lambda$ should be
even non-negative integer or $\lambda =1$, corresponding to the fundamental
representation in this case. In the case $so(3)$ also half-integer values of
$-\lambda$ fulfil the finiteness conditions.
The example of the Jordan-Schwinger $L$ operator considered below has just
this form of weights.

In analogous way considering the case of the vanishing of the third factor
for other values of $i=i_0 $
we obtain the solution with the sequence of weights
$ \lambda_i^{[1]}$ of the form
\be \label{lambda0}
(\lambda, \ldots,\lambda, 0, \ldots,0). \ee 
The first entry $0$ stands at the position $i_0+1$.
Now the value of $k$ is fixed at
$$k = - \half (\beta-i_0)^2 - \half \lambda^2 + \epsilon (\beta-i_0+1) \lambda. $$
If the rank $m$ exceeds $i_0+1$ then the weight function ratios are
$$  1, ...,1, \frac{u-\half \epsilon \lambda -\half (\beta-i_0)}
{u+\half \epsilon \lambda -\half (\beta-i_0)}, 1, \ldots, 1 $$
with the non-trivial entry at the position $i_0$. The modifications for the
ranks $m= i_0$ and $m=i_0+1$ are analogous to the ones at $i_0=1 $.

Let us consider the case that the third factor in (\ref{3 factors}) 
vanishes for $i=1 $ and $i=2$,
 and solve for $\lambda_1, \lambda_2, \lambda_3$.
$$ k + \half (\beta-1)^2 + \half (\lambda_1^2 + \lambda_2^2) - \epsilon \beta \lambda_1
- \epsilon(\beta-1) \lambda_2 = 0, $$
$$ k + \half (\beta-2)^2 + \half (\lambda_2^2 + \lambda_3^2) - \epsilon  \lambda_1
- \epsilon(\beta-1) \lambda_2 -  \epsilon(\beta-2) \lambda_3 = 0. 
$$
If we do not fix the value of $k$ beforehand, it will be fixed by
choosing two of the three weights. 
The first condition is fulfilled by fixing $k$, 
\be \label{k12}
 k= - \half (\beta-1)^2 - \half \lambda_1^2 + \epsilon \beta \lambda_1 
- \half \lambda_2^2 + \epsilon (\beta-1) \lambda_2, \ee
but the consistency of both
conditions implies the following restriction on $\lambda_1$ and $\lambda_3$ 
\be \label{lambda13}
 (\lambda_1- \epsilon)\big(\lambda_1 + \epsilon - 2\epsilon (\beta -1)\big) = 
\lambda_3 \big(\lambda_3- 2\epsilon(\beta-2)\big),
 \ee
with arbitrary values of $\lambda_2 $.
We calculate the first weight function ratio using (\ref{k12}),
$$ f_1(u) = \frac{u + \frac{\epsilon}{2}(\lambda_2-\lambda_1) -\half
(\beta-1)}{u - \frac{\epsilon}{2}(\lambda_2-\lambda_1) -\half (\beta-1)},
$$ 
and the second using  (\ref{k12}),  (\ref{lambda13}),
$$ f_2(u) = \frac{(u + \frac{\epsilon}{2}(\lambda_2+\lambda_1) +\half
(\beta-1)) (u - \frac{\epsilon}{2}(\lambda_2-\lambda_1) -\half
(\beta-1)}{(u + \frac{\epsilon}{2}(\lambda_3+\lambda_2) +\half
(\beta-2)) (u - \frac{\epsilon}{2}(\lambda_3-\lambda_2) -\half
(\beta-2)}.
$$
Here $\lambda_1$ and $\lambda_3$ are related by (\ref{lambda13}).
Let us proceed with its  particular solution $ \lambda_1 - \lambda_3 =
\epsilon$, then we obtain $ f_2(u) = 1$.
In this case, for $so(2m)$ and $sp(2m)$ $m>2$, 
we have the sequence of allowed weights
$$( \lambda_1, \lambda_2, \lambda_1-\epsilon, ..., \lambda_1-\epsilon ) $$ 
 and the  sequence of weight function ratios
\be \label{lambda1lambda2}
\frac{u + \frac{\epsilon}{2}(\lambda_2-\lambda_1) -\half
(\beta-1)}{u - \frac{\epsilon}{2}(\lambda_2-\lambda_1) -\half (\beta-1)},
1, \ldots, 1.
\ee 
The straightforward generalization of this solution  is given by 
the sequence of allowed weights
$$ (\lambda_1, \ldots, \lambda_1,\lambda_2, \lambda_1-\epsilon, \ldots,
\lambda_1-\epsilon) $$
and the  sequence of weight function ratios
\be \label{lambda1ldots}
 1, \ldots, 1, \frac{u + \frac{\epsilon}{2}(\lambda_2-\lambda_1) -\half
(\beta-i_0+1)}{u - \frac{\epsilon}{2}(\lambda_2-\lambda_1) -\half
(\beta-i_0+1)}, 1, \ldots, 1,
\ee
where $\lambda_2$ and the non-trivial ratio are standing at the position
$i_0$ and $m>i_0+2$ is assumed. 

In the case $so(2m+1), m>2$ the forms (\ref{lambda0}) of the weight
sequences are allowed too. (\ref{lambda1ldots}) holds for the 
particular value $\lambda_1= \epsilon=1$.

\section{Examples of the linear evaluation}
\setcounter{equation}{0}

\subsection{Presentations based on Clifford algebras}

 For the spinorial representations \cite{Re,SW,CDI1,IKK15} 
let us consider  the case that the generators are composed in terms
of an underlying algebra, which in turn is generated by $c^a$,
obeying the commutation relations of quantum oscillators or of the Clifford algebra
 \be\label{osc}
c^ac^b+\epsilon c^bc^a=\e^{ba},\qquad c_a=\e_{ab}c^b,\qquad
\Rightarrow\qquad c_ac^b+\epsilon c^bc_a=\d^b_a,
 \ee
$$
c_ac_b+\epsilon c_bc_a=\epsilon\e_{ab},\qquad\qquad c_ac^a=
\frac n2=\epsilon c^bc_b,
$$
We recall that $n$ is the dimension of the fundamental representation space
$V_f$.
We may consider $c^a$ as operators in the spinor space  (of dimension $2^m$)
or in  the Fock space of  fermions or bosons.

Let us compose the Lie algebra generators as
 \be\label{Gosc}
G_{ab}=\frac\epsilon2 \e_{ab}-c_ac_b,
\ee
and $L(u)$ as 
$$L_{ab}(u) = u \e_{ab} - \half (c_a c_b - c_b c_a) = (u+ \half ) \e_{ab} - c_a
c_b. $$

We check that the Lie algebra relations are fulfilled.
Further,
$$ (G^2+\b G)_{ab}=\frac\epsilon4(n-\epsilon)\e_{ab}=
\frac\epsilon2(\b+\frac\epsilon2)\e_{ab}.
 $$
In this way the composite generators (\ref{Gosc}) fulfil the additional 
condition of the linear evaluation.
Thus the spinor representation (\ref{osc})
provides an example of the linear evaluation of the Yangian $
\mathcal{Y}^{(1)}({\cal G})$.

The particular metric choice (\ref{metric}) and the index range separation
divides also the set $c^a$ into rising and lowering operators.
In the case of $so(2m+1)$ we have additionally the idempotent 
operator $c_0$, $\;\;2 c_0^2 = 1$.
A vector $|0\rangle$ obeys the highest weight conditions (\ref{hwY}),
(\ref{hwY0}) if
$$
c_{-i} |0\rangle= 0, \qquad i=1,\ldots, m.
$$
We calculate the weights from
$$
G_{-ii} |0\rangle= \lambda_i^{[1]} |0\rangle,\qquad G_{-ii} = 
\half -  c_{-i} c_i.
$$
The weights of the representation built on  this vector are
\be \label{-half}
( \lambda_1, \ldots, \lambda_m)= (-  \half,\ldots, - \half). \ee

We have additionally the highest weight vector obeying in the $so(2m)$
case
$$
c_{-i} |\tilde 0\rangle= 0, \ i=1,\ldots, m-1, \qquad
 c_{m}  |\tilde 0\rangle= 0,
 $$
with the weights 
$$ (-  \half,\ldots, - \half, + \half). $$

We notice that $|\tilde 0\rangle= c_{m} |0\rangle$ and find that also 
in the
symplectic case the analogous vector obeys the highest weight conditions.
The weights are 
$$ (-  \half,\ldots, - \half, - \frac{3}{2}). $$
We check that the weight conditions  (\ref{lambdalin}) are fulfilled.
In the case of $|0\rangle$ the condition is fulfilled because the first factor
in (\ref{lambdalin}) 
vanishes. In the case of  $|\tilde 0\rangle$ the change in the value 
of the weight
$ \lambda_{m}$ is related to the second factor (\ref{lambdalin}), 
vanishing with this value at $i=m-1$.

The ratios of the weight functions defined as (\ref{lamblin1}), 
$ \lambda_i(u) = -u\epsilon + \lambda_i, 
\qquad \lambda_{-i}(u) = -u -\epsilon  \lambda_i
$,
are in the $so(2m)$ representation built on $|0\rangle$ a particular case of
(\ref{lambdadots})
$$  1, \ldots, 1,  \;  \frac{u+\half}{u-\half},  
$$
and in the $so(2m)$ representation built on $|\tilde 0\rangle$ a particular case
of (\ref{lambdadots-})
$$  1,  \ldots, 1,  \;   \frac{u+\half}{u-\half}, \; 1 .$$
The finiteness criterion is fulfilled in both cases. The sequence of Drinfeld
polynomials $ P_1(u), \ldots, P_m(u) $
is in the case $|0\rangle$
$  1 , \ldots, 1,  u-\half, $
and in the case $|\tilde 0\rangle$
$  1 , \ldots, 1,  u-\half, 1 $.
The ratios of the weight functions are
  in the $sp(2m)$ representation built on $|0\rangle$ a particular case of
(\ref{lambdadots})
$$ 1, \ldots,1 ,  \frac{u-\half}{u+\half}, $$
and in the $sp(2m)$ representation built on $|\tilde 0\rangle$ a particular
case of (\ref{lambdadots-2})
$$  1,  \ldots, 1,    \frac{u-\half}{u-\frac{3}{2}},
\frac{u-\frac{3}{2}}{u+\frac{3}{2}}. $$
The finiteness criterion is not fulfilled. 

%----------------------------------------------------------------

In the case of $so(2m+1) $ we have additionally the highest weight condition
(\ref{hwY0}), which is fulfilled by $|0\rangle$ but not by $|\tilde 0\rangle$.
(\ref{-half}) is the only  weight sequence here. 
The weight function ratios  in the $so(2m+1)$ representation built on $|0\rangle$ 
are a particular case of (\ref{-half0})
$$  1, \ldots, 1,  \;  \frac{u+\half}{u}. $$
The finiteness criterion is fulfilled and the Drinfeld polynomials are
$ 1,\ldots, 1, \;  u $.

In sect. 5.2 we have seen that 
there is no continuous parameter in the allowed sequence of weights for the
linear evaluation of $so(2m+1)$ type Yangian representations.
The described  spinorial representation and a finite number  of related
representations are the only ones in the linear evaluation. 

In the cases $so(2m)$ and $sp(2m)$ there are other representations of the
linear evaluation depending on one continuous parameter.

\subsection{Presentations  based on  Heisenberg algebras}

In the cases  $so(2m), sp(2m)$
there are examples of linear evaluation representations
constructed in \cite{RF20} for the case $so(2m)$. 

The matrix of generators $ G = (G^a_b)$ results in the $L$ operator of
the linear evaluation 
$ L(u) = I u +  G $
if  besides of the Lie algebra relations (\ref{Liealg0}) the constraint
\be \label{c13}
  G^2 +  \beta  G = c_2 I, 
  \ee 
is fulfilled (compare \ref{symC1368} for $H=0$).

A similar relation holds for the matrix of generators $G^{s\ell}$ 
of the Jordan-Schwinger presentation of $s\ell(2)$,
$$(G^{s\ell})^2 +  G^{s\ell} = \ell(\ell+1) I. $$
It is connected with the property of factorization \cite{SD05}
$$ G^{s\ell} =
\begin{pmatrix}
-\ell -1 +\dd x & \dd \\
x(2\ell+1 - \dd x) & \ell - x\dd
\end{pmatrix}
= 
\begin{pmatrix}
1 & 0 \\
-x & 1
\end{pmatrix}
\begin{pmatrix}
- \ell -1 & \dd \\
0 & \ell
\end{pmatrix}
\begin{pmatrix}
1 & 0 \\
x & 1
\end{pmatrix}.
$$
The $L$ operator $L^{s\ell}(u) = I u + G^{s\ell}$ obeys the RLL equation
(\ref{rll}) with Yang's $R$ matrix $R (u) = I u + P$, where the matrix
elements of $I$ and $P$  are as in (\ref{IKP})
 with the indices taking the values $1,2$ only. 
 
For the cases of $so(2m)$ and $sp(2m)$ we construct the generator
matrix $G$  of size $2m \times 2m$ by replacing the single Heisenberg pair
$x, \dd $ by $\half m (m \mp 1)$ pairs $x^i_j, \dd^k_l$, antisymmetric or
symmetric in the indices.  In order to obey (\ref{c13})
we copy the above form with
$x, \dd$ replaced by $m \times m$ matrices $\hat x, \hat \dd$ with 
 matrix elements $x^i_j, \dd^k_l$. They are assumed 
to be antisymmetric in the orthogonal
and symmetric in the symplectic case, 
$$ x^i_j = - \epsilon x^j_i, \qquad \dd^k_l = -\epsilon \dd^l_k.  $$
The entries without $x$ or $\dd$ are
replaced as proportional to the $m\times m$ unit matrix $\hat I$.
The constraint on $G^2$ is fulfilled by the factorized form if we modify
the first block on the diagonal  by replacing $-\ell-1$ by $-\ell-\beta$
$$  G = 
\begin{pmatrix}
\hat I & 0 \\
-\hat x & \hat I
\end{pmatrix}
\begin{pmatrix}
- \ell \hat I -\beta \hat I & \hat \dd \\
0 & \ell
\end{pmatrix}
\begin{pmatrix}
\hat I & 0 \\
\hat x & \hat I
\end{pmatrix}.
$$
(\ref{c13}) holds with $c_2= \ell (\ell+\beta)$.

We assume the commutation relations of Heisenberg type,
\be \label{ddijxkl}
 [\dd^i_j, x^k_l] = \delta^k_j \delta^i_l - \epsilon \delta^i_k \delta^j_l, \ee
 compatible with the $\epsilon$ anti-symmetry of the matrices.

The reason for the replacement $1 \to \beta$ is seen from the commutation
relations,
$$ \dd^i_k x^k_j  =  \delta^k_k \delta^i_j - \epsilon \delta^i_k \delta^k_j 
  + x^j_k \dd^k_i = \beta \delta^i_j + x^j_k \dd^k_i . $$
The matrix of generators appears in a $2 \times 2 $ block form with 
$m \times m$ matrices in the blocks.
The matrix elements are labelled 
by signs  $\sigma, \rho= \pm$ referring to the blocks  and by 
$i,j=1, ..., m$. 
$$ G^{+ i}_{ \ + j} = - (\ell +\beta) \delta^i_j + \dd^i_k x^k_j, $$
$$ G^{+ i}_{ \ - j} =   \dd^i_j, $$
$$ G^{- i}_{ \ + j} = x^i_k ((2\ell+\beta) \delta^k_j - \dd^k_l x^{l}_j ), $$
$$G^{- i}_{ \ - j} = \ell \delta^i_j - x^i_k \dd^k_j. $$

We adopt the metric (\ref{metric}) and use it to lower
indices.

$$ G_{-i, + j} = - \epsilon (\ell +\beta) \delta^i_j + \epsilon \dd^i_k
x^k_j, $$
$$ G_{-i,  - j} =  \epsilon \dd^i_j, $$
$$ G_{ +i,\ + j} = x^i_k ((2\ell+\beta) \delta^k_j - \dd^k_l x^{l}_j ), $$
\be \label{Gell}
 G_{+i, \ - j} = \ell \delta^i_j - x^i_k \dd^k_j .
 \ee

The $\epsilon$ anti-symmetry of $G, \ G_{ab} = - \epsilon G_{ba}$ holds.

As the main point the commutation relations (\ref{Liealg0}) have to be
checked. 

$$ [G_{-i_1,+j_1},G_{-i_2,j_2 } ] = [(\dd x)_{i_1, j_1}, (\dd x)_{i_2, j_2}]
= 
- \delta^{i_1}_{j_2} (\dd x)^{i_1}_{j_2} + \delta^{i_1}_{j_2} (\dd
x)^{i_2}_{j_1} + \epsilon \dd^{i_1}_{i_2} x^{j_1}_{j_2} - \epsilon \dd^{i_2}_{i_1}
x^{j_2}_{j_1} = 
$$
$$
- \epsilon \delta^{i_1}_{j_2}G_{-i_1, +j_2} + \epsilon \delta^{i_1}_{j_2}
G_{-i_2,+j_1},
$$

$$ [G_{-i_1,+j_1},G_{-i_2,-j_2 } ] = [(\dd x)_{i_1, j_1}, \dd^{i_2}_{j_2} ]
= 
- \dd^{i_1}_{j_2} \delta^{i_2}_{j_1} + \epsilon \dd^{i_1}_{i_2} \delta^{j_1}_{j_2} =
- \epsilon G_{-i_1, -j_2} \delta^{i_2}_{j_1} - \epsilon G_{- i_2, - i_1}
\delta^{j_1}_{j_2}. $$
Here the $\epsilon$ anti-symmetry has been used. 
$$ [G_{+i_1,+j_1},G_{-i_2,-j_2 } ] = (2 \ell + \beta) \epsilon [x^{i_1}_{j_1},
\dd^{i_2}_{j_2} ] - \epsilon [ (x\dd x)^{i_1}_{j_1}, \dd^{i_2}_{j_2} ] =
(2 \ell + \beta) \epsilon (- \delta^{i_1}_{j_2} \delta^{i_2}_{j_1} + 
\delta^{i_1}_{i_2} \delta^{j_2}_{j_1} ) + $$ $$
 + \epsilon \{ (x\dd)^{i_1}_{j_2} \delta^{i_2}_{j_1} - \epsilon (x\dd)^{i_1}_{i_2}
\delta^{j_1}_{j_2} + \delta^{i_1}_{j_2} (\dd x)^{i_2}_{j_1}   
 - \epsilon (\dd x)^{j_2}_{j_1} \delta^{i_1}_{i_2} \} = $$ $$
 - \epsilon \delta ^{j_1}_{i_2} G_{+i_1,-j_2} + \delta^{i_1}_{j_2} G_{-i_2, j_1} 
+ \delta^{i_1}_{i_2} G_{+j_1,-j_2} - \epsilon \delta^{j_2}_{j_1} G_{-i_2,i_1} .$$
We calculate the weight functions of the representations  
by polynomials 
of the matrix elements $x^i_k$ and the highest weight vector $1$. 
$$ L_{ab} = u \e_{ab} + G_{ab}, $$
$$ L_{-i,+i}(u) \cdot 1 = \lambda_i(-u) \cdot 1, \qquad    
L_{+i,-i}(u) \cdot 1 = \lambda_{-i}(-u) \cdot 1, $$
$$ \lambda_i (u) = -\epsilon (u+\ell), \qquad   \lambda_{-i}(u) = -u+\ell. $$
For the weight function ratios we obtain
\be \label{u+lu-l} 
\frac{\lambda_i (u)}{\lambda_{i+1} (u) }= 1, \qquad   i=1, \ldots, m-1, \qquad  
 \frac{\lambda_m (u)}{\lambda_{-m} (u) }= 
\frac{\lambda_m (u)}{\lambda_{1-m} (u) }
= \epsilon \frac{u+\ell}{u-\ell}. 
\ee
The sequence of weight function ratios is in both $so(2m) $ and $sp(2m)$
cases
$$ 
 1, \ldots, 1,  \frac{u+\ell}{u-\ell}. 
$$ 
It coincides with (\ref{lambdadots}) for $\lambda= -\epsilon \ell$. 

For non-negative integer values of $2\ell$ 
the only non-trivial ratio $f_m(u)$ can be written
in terms of the polynomial  
$ P(u) = (u-\ell) (u-\ell +1) \ldots(u+\ell-1) $ as
$ \frac{P(u+1)}{P(u)}$. In the orthogonal case this is the condition
for the representation module  to have a finite-dimensional sub-module (\ref{Pi}).

In the symplectic case the finiteness condition (\ref{Pi})
is however
 $$ \frac{\lambda_m (u)}{\lambda_{-m} (u) }= -
 \frac{P(u+2)}{P(u)} $$
We see that $2\ell$ should be an even non-negative integer 
to obey this condition with the polynomial 
$ P(u) = (u-\ell) (u-\ell +2) \ldots(u+\ell-2) $.

By comparing the particular case of $sp(2)$ with $s\ell(2)$
we understand
why the condition for terminating the sequence of excitations by
 action by $(G_{+1,+1})^r$ on $|0\rangle= 1$ can be fulfilled by some value of 
 $r$ only if $2\ell$ is an even integer. 
Indeed, whereas in the case of $s\ell(2)$
the commutation relation of the basic Heisenberg pair is
$[\dd, x] = 1 $,
 the modified commutation relation (\ref{ddijxkl})
used in the construction
implies in the case $sp(2)$, where the indices $i,j$ take only a single value,
$[\dd, x] = 1+1 $.

\section{Examples of the quadratic evaluation}
\setcounter{equation}{0}

\subsection{Product of linear $L$}

We consider $L$ operators constructed by  product of
two $L$ operators of the linear evaluation.
$$ 
L_1(u) = I_1 u +G_1, \qquad L_1(u) = u I_2 + G_2,\qquad  G_i^2 + \beta 
G_i =c_2^{(i)} I_i, $$
$$ L_{12}(u) = L_1(u) L_2(u+\delta). $$
We obtain
$$ L_{12}(u) = I u^2 + u(G_1+G_2+ \delta I) +  \delta G_1 +
G_1 G_2 =
I u^{\prime \ 2} + u\p G + H,\qquad  u\p = u - \delta\p. $$
We choose $\delta\p = - \half \delta$, such that
$$ 
G= G_1+G_2. 
$$
We obtain for the  $\epsilon$-antisymmetric parts of
$H$,
$ H= \half(G^2 + \beta G + k) + \bar H  $, 
$$ 
\bar H= \half (G_1G_2-G_2G_1) + \half \delta (G_1-G_2) , \ \ k= - \half \delta^2
- c_2^{(1)}-c_2^{(2)}. $$
The resulting $L_{12}$  depends on $\delta$ and on the parameters in the factors.
In the cases $so(2m)$ and $sp(2m) $ we have $L$ operators depending on one
parameter (\ref{c13}), (\ref{Gell}). If we choose  as $L_1, L_2$ such
expressions the resulting $L_{12}(u)$ depends on
three parameters $\ell_1, \ell_2, \delta$.

Let the representations of the factors be built on the corresponding highest
weight
vectors $|0\rangle_1, |0\rangle_2 $ and consider the representation
of $L_{12}(u)$ built on $|0\rangle_1 \otimes |0\rangle_2 $.

The weight function ratios multiply,
$ f_{12, i}(u) = f_{1, i}(u) f_{2, i}(u+ \delta), $
the resulting sequence of ratios is
$$  1, \ldots, 1, \  \ \frac{(u+\ell_1)(u+ \delta +\ell_2)}{(u-\ell_1)(u+ \delta-\ell_2)}. $$
The parameters enter the weight functions in the following way.
$$\lambda_i = -\e(\ell_1+ \ell_2), \quad \bar \lambda_i^{[2]} = -\frac{\e
\delta}{2} (\ell_1-\ell_2), $$ 
$$ k= -\half \delta^2 - \ell_1(\ell_1+\beta) - \ell_2(\ell_2+\beta) =
-\half \delta^2 - \half(\ell_1 +\ell_2 +2\beta) (\ell_1+\ell_2) -
\half(\ell_1 - \ell_2)^2. $$

We have the case of non-vanishing $\bar \lambda_i^{[2]}$. These 
weight components as well as $\lambda_i^{[1]}$ are independent of $i$ and
therefore   (\ref{u1i}) holds. We have a particular case of
(\ref{lambdabarlambda}).

With the $L$ operators constructed on Clifford algebras (\ref{Gosc})
we obtain other cases of non-vanishing  $\bar \lambda_i^{[2]}= \bar
\lambda_i$.
Let us  choose both $L_1$ and $ L_2$ in the form (\ref{Gosc}) and consider the
representation built on $ |0\rangle \otimes |\tilde 0\rangle$. 
Actually, only  the $m$th component of $ \bar \lambda_i$ is
non-zero, $\bar \lambda_m \not = 0$. (\ref{u1i}) is fulfilled
for $i=1, ..., m-2$ because here the weight components do not depend on $i$.
(\ref{u1i})  for $i=m-1$ holds in the way that the expressions on both sides
vanish in both the $so(2m)$ and $sp(2m)$ cases.    We obtain for the weight
sequences $(\lambda_1, ..., \lambda_m); (\bar
\lambda_1, ..., \bar \lambda_m) $ 
in the orthogonal case
$$(-1,..., -1,0);  \ (0, ..., 0, \frac{\delta}{2} ) $$
and in the symplectic case
$$(-1,..., -1,-2); \  (0, ..., 0, \frac{\delta}{2} ). $$
The weight function ratios are calculated as above by multiplication.

\subsection{Jordan-Schwinger presentations}

We consider the quadratic evaluation
(\ref{LGH})
in the case where the $\e$ anti-symmetric part of $H$ vanishes,
then by  (\ref{symC1368})
\be \label{H}
H= \half (G^2 + \beta G + k I). 
\ee
$k$ is the value of $c^{(2,3)}$  (\ref{symC1368}) in the present case.

In the prominent example of Jordan-Schwinger  
the generators $G$ are built from the canonical pairs
\be \label{ddx}
 [\dd_a, x_b]_{-\epsilon} = \e_{ab} ,
 \ee
in the following way
\be \label{GJS}
  G_{ab} = -\epsilon( x_a \dd_b -\epsilon x_b \dd_a).
\ee
The index range includes $n$ values, as above we choose 
$a,b = -m, \ldots, -1, (0), +1, \ldots m$ with the value $0$ included in the case
of $so(2m+1)$.

The matrix elements of $G$ obey (\ref{Liealg0}).
For the matrix elements of $H$ we obtain
\be \label{2H}
 2H_{ab} = 
\big((x\dd)+(\beta-1)\big) \epsilon (x_a \dd_b+\epsilon x_b \dd_a)+\big((x\dd) 
+k\big)\e_{ab} - x^2 \dd_a \dd_b - x_a x_b x^2 . 
\ee
$(x\dd)$ denotes the scalar product, $ (x\dd)= x_a \dd^a$. 
 
The quadratic Casimir operator is obtained as
\be \label{c2}
nc_2= Sp G^2=\big(G^2\big)^a_{ \ \ a}=2 \epsilon \{ (x\dd)^2-x^2 \dd^2+
(1+ \epsilon)\beta (x\dd) \}.
\ee
Here $x^2 = x_a x^a,\;\; \dd^2 = \dd_a \dd^a$.

We consider representations by functions of $x_a$.
The highest weight conditions in the form  (\ref{hwY}),(\ref{hwY0})
are fulfilled with the highest weight  vector represented by 
\be \label{psi}
 \psi(x) = (x_{-1})^{2\ell}. 
 \ee
The representation of functions  of $x_a$  built on $\psi$  is spanned by
the homogeneous functions of degree $2\ell$. Therefore
in the above expression the dilatation operator $(x\dd)$ is replaced by
$2\ell$. Studying the action of the excitation elements in $G$ we see that
for integer values of $2\ell$ the representation space  has a 
finite-dimensional  subspace invariant under the Lie algebra action.

We notice that in the symplectic case, where the canonical pairs are of
Grassmann type, the value of the representation parameter is restricted to
$2\ell= 1$ or $ 2\ell=0$. The latter results in the trivial 
and the former in the fundamental representations. It results also in 
an apparent ambiguity in the $\epsilon $ dependence of the above
expressions (\ref{2H}), (\ref{c2}) due to relations like
$ (1-\epsilon ) \big((x\dd) -1\big) x_a \dd_b = 0 $.

%The weight functions defined as eigenvalues of $L_{-i,i}(u)$ in action of $\psi(x)$
%as in (\ref{hwY}) have the form
%$$\lambda_i(-u) = \epsilon u^2 + u \lambda^{[1]}_i +  \lambda^{[2]}_i .
%$$
%The eigenvalue of $G_{-i,i}$ in the action on $\psi(x)$ is $ \lambda^{[1]}_i$, 
%and
%the eigenvalue of $H_{-i,i}$ in the action on $\psi(x)$ is $ \lambda^{[2]}_i$.

We investigate the $g\ell(2)$ Yangian subalgebra $\mathcal{Y}^{-1,2}$ with
index values $a= -i=-1,-2,$ $b=j=1,2$ and consider the representations $D$ on
functions of $x_{-1}, x_{-2} $.

$$
(L_{-i,j})|_{{D}} =u^2 \e_{-i,j} + u (G_{-i,j})|_{{D}} + (H_{-i,j})|_{{D}}, 
$$
$$ 
(G_{-i,j})|_{{D}} = -\epsilon x_{-i} \dd_j, \qquad2 (H_{-i,j})|_{{D}} = \big((x\dd) -1 
+ \beta\big) \epsilon x_{-i} \dd_j + \big((x\dd) + k\big)\e_{-i,j} - 0. 
$$
On functions homogeneous of degree $2\ell$ we substitute $(x\dd) \to
2\ell $ and obtain
$$ 
L_{-i,j}(u)|_{D,\lambda}=\big(u^2 + \half(2\ell+k)\big) \e_{-i,j}+\half (2\ell-1+
\beta  -2 u )\epsilon  x_{-i} \dd_j. 
$$
The reduced monodromy should be transformed to the standard JS form of
$g\ell(2)$ $L$ operators,
$L^{g\ell} (u) = I u - x_a\dd_b$, 
by allowed equivalence operations. 
This is possible if the first coefficient, quadratic in $u$, can be divided
by the second, linear in $u$. $k$ is fixed by this condition
\be \label{k}
 k = - (x\dd) - \half (\beta  + (x\dd) -1)^2  \ee
up to contributions vanishing in the considered representation.

Representations built on the highest weight vector $\psi = x_{-1}^{2\ell} $ result in
the weight function ratio $f_1(u)$. 
For other ratios of weight  functions we have to go beyond the subalgebra
$\mathcal{Y}^{-1,2}$ of $g\ell(2)$ type. We calculate the weight functions from  
$$G_{-1,1} \psi = -\epsilon 2\ell\psi , \qquad G_{+1,-1} \psi = 2\ell \psi, 
\qquad G_{-i,i} \psi = 0,\qquad
 G_{i,-i} \psi = 0,\qquad i>1 $$
and 
$$2 H_{-1,1}\psi = \epsilon 2H_{+1,-1}\psi= [(2\ell+k) \epsilon +
(2\ell-1+\beta) \epsilon 2\ell
]\psi, $$
$$2 H_{-i,i}\psi = \epsilon 2H_{+i,-i}= (2\ell+k) \epsilon \psi,  \qquad i>1. $$

The sequence of weights coincides with (\ref{lambda00}) with the substitution
$\lambda=-\epsilon 2\ell$. We check now by direct calculation that also the 
weight function ratios coincide.

We substitute (\ref{k}),
$ 2\ell + k = - \half (\beta  -1+2\ell)^2 $, 
and obtain the weight functions
$$ \lambda_i(-u) = u^2 \epsilon - \frac{1}{4}  (\beta \epsilon -1+2\ell)^2,  \qquad i\not = 1 
$$
$$ \lambda_1(-u) = u^2 \epsilon - u 2\ell \epsilon + \half \{ 2\ell (2\ell-1 +
\beta) \epsilon
- \half \epsilon  (2\ell-1 + \beta )^2 \} = $$ $$
\epsilon (u- \half (2\ell + \beta-1)) (u- 2 \ell + \half  (2\ell + \beta-1))  $$
We obtain also
$$ \lambda_{-1}(-u) = \epsilon  \lambda_1(+u) . $$

We obtain the sequence of weight function ratios 
$$ f_1(u), 1, ..., 1, $$ 
\be \label{f1JS} 
f_1(u) = \frac{u+2\ell -\half (2\ell + \beta-1)}{u -\half (2\ell +
\beta-1)}. 
\ee
This applies for $so(2m+1)$ for all ranks $m$ and for $so(2m)$ for $m>2$
with arbitrary values of $2\ell$. It applies for $sp(2m)$ where by
construction $2\ell$ is restricted to the values $0$ and $1$.
In the case $so(4)$ we have the two weight function ratios coinciding
$f_1(u) = f_2(u)$   with $f_1(u)$ given in (\ref{f1JS}) and $\beta = 1$.

In this way we have confirmed that the the representations resulting from
the Jordan-Schwinger $L$ operator constructed with $G$ as in (\ref{GJS}) on
the highest weight vector represented by (\ref{psi}) covers the
 the quadratic evaluation representations (\ref{lambda00}),
(\ref{flambda00}) for all
values of $\lambda$ in the orthogonal cases and for the distinguished values
$\lambda=0, 1$ for the symplectic case.

\subsection{$L$ operators of spinorial type}

We consider further cases of the quadratic evaluation (\ref{LGH}) where
the $\epsilon$-antisymmetric part of $H$ vanishes (\ref{H}).
Now $G$ is not constructed like in (\ref{GJS}). We consider the elements of 
$\mathcal{G} \otimes \mathcal{G}$  
\be \label{W}
 W_{ab,cd} = G_{ab} G_{cd} + G_{ac} G_{db} + G_{ad} G_{bc} +
G_{cd} G_{ab} + G_{db} G_{ac} + G_{bc} G_{ad} \ee
$$
=2(G_{ab}+\epsilon\e_{ab})G_{cd}+2(G_{ac}+\epsilon\e_{ac})G_{db}
+2(G_{ad}+\epsilon\e_{ad})G_{bc}=0.
$$
and impose the condition of vanishing for all index values \cite{KK17}. 
 This condition implies in particular the vanishing of the third order
polynomial in $G$,

 \be\label{chi3}
W_{ab,cd}G^{bc}=4\chi^{(3)}_{ad}=0,\qquad
\chi^{(3)}(G) = G^3+(\epsilon+2\b) G^2+(2\epsilon\b-{
\epsilon n c_2}) G- {n c_2}.
 \ee
We stress that the particular representation for $G$
is defined by the algebraic restriction (\ref{chi3}).

The condition is fulfilled for the Jordan-Schwinger form (\ref{GJS})
of the generators and in particular for the fundamental representation.

The vanishing condition for $W$ (\ref{W}) is connected to the linear 
spinorial evaluation,
$$ \mathcal{L}(u) = I u + \hat G, \quad \hat G = -\half c^a c^b G_{ab}. $$
$c^a$ are the generators of the Clifford algebra (\ref{osc}).
If (\ref{W}) vanish for $G_{ab} $ then $\mathcal{L}(u)$ obeys the
modification of the Yang-Baxter RLL-relation (\ref{rll}) with the
fundamental $R$ matrix replaced by the spinorial $R$ matrix
\cite{SW,IKK15,KK19}. The condition (\ref{W}) can be written in a form
analogous to the condition of linear evaluation (\ref{c13}),
$$ \hat G^2 + \beta \hat G = \half c_2 I. $$

 Consider  the action of $W$ (\ref{W})
on the highest weight vector, in particular of its tensor components
 $ab,cd = -i i, -j j, \quad i<j$,
$$
W_{-i i,-j j} |0\rangle= \lambda_i \lambda_j + [G_{-i-j}, G_{ji}] |0\rangle+ G_{-ij},G_{i-j}]
|0\rangle+ \lambda_i \lambda_j + 0 + 0 =
$$
$$
=2 \lambda_i \lambda_j -\lambda_i - \lambda_j + \lambda_i - \lambda_i = 2 (\lambda_i-1) \lambda_j,
$$
and $ab,cd = -i i, -i i$,
$$
W_{-ii,-ii} |0\rangle= 2 \lambda_i^2 + [G_{-i-i}, G_{ii}] |0\rangle-2\e \lambda_i^2 =
2(1-\e) (\lambda_i - 1) \lambda_i.
$$
The condition $W= 0$ implies 
\be \label{hi2}
  (\lambda_i-1) \lambda_j = 0, \quad (i<j), \qquad (1-\e) (\lambda_i - 1) \lambda_i= 0.
\ee
The second relation does not imply any restriction on $\lambda_i$ in the
orthogonal case where $\e= +1$. In the symplectic case it restricts
the values of $\lambda_i$ to be either $0$ or $1$. The first relation is fulfilled
in particular if $\lambda_i = 0,\; i>1$ as in the Jordan-Schwinger example
and in general by weight sequences of the form
$$ (1, \ldots, 1, \lambda, 0, \ldots, 0). $$
In the orthogonal case this class of representations coincides with
(\ref{lambda1ldots}) for the particular values $\lambda_1 = 1$ and
$\lambda_2= \lambda$. In the symplectic case we
notice the coincidence with (\ref{lambda0}) with the particular 
value $\lambda=1$.

\section{Discussion}

The extended Yangian algebra of the orthogonal or symplectic types
has a non-trivial center. Factorizing the center leads from the extended
Yangian algebra as defined by the Yang-Baxter $RLL$ relations to the 
Yangian algebra according to Drinfeld definitions
\cite{Drinfeld,Drin,Drin88,JLM17,GRW17}. 

The algebra relations of the extended Yangian algebra
$\mathcal{Y}(\mathcal{G})$
appear from the $RLL$ relations as equalities in 
$End (V_{f})_1\otimes End (V_{f})_2 \otimes \mathcal{Y}$, where $V_f$ is the
fundamental representation space. The part of these relations symmetric in
the permutation of the first two factors can be obtained more directly from the 
generating function $C(u)$ of the center. 

The highest weight representations can be described by the weight functions
$\lambda_a(u)$. In our conventions $\lambda_a(u)$ is the eigenvalue of 
$L_{-a,a}(-u)$ in action on the highest weight vector $|0\rangle$. 
 A sequence of embedded Yangian representations of the same type but lower
rank is a powerful tool in studying representations \cite{AMR05}.  

Representations
can be characterized by $m$ ratios of weight functions chosen as in
(\ref{fi}) related to the simple roots. Yangian representations of
$g\ell(2)$ type can be associated to each of these ratios. For most ratios
 $f_i(u), i=1, ..., m-1$,  this association 
can be done in a straightforward way by choosing sets of matrix elements of
$L(u)$ as generators of the corresponding $g\ell(2)$ Yangian. 
It works less trivially for
$f_m(u)$ leading in the cases of $sp(2m)$ and $so(2m+1)$ to changes in the
weight function arguments. 

It is known that these ratios turn out to be ratios of polynomials in the
spectral parameter with definite shifts of arguments in the case of
finite-dimensional representations \cite{VOT84,Drin88,AMR05}. 
The conditions of finiteness 
can be formulated as relations between the zeros and poles of the ratio
functions $f_i(u)$. 

Relations between the weight functions and their expansion components are
derived from the center generating function $C(u)$. We have recovered in
this way the general
relations derived in \cite{AMR05}. We have formulated the relations for the
cases of linear and quadratic evaluation and have studied their solutions.  
There are $m-1$ such relations restricting the weight functions in all cases
and an additional $m$th relation in the $so(2m+1)$ case. 

The conditions on the weight components of the linear evaluation
representations factorize in two factors linear in the weights. 
In the cases $sp(2m)$ and $so(2m)$ there are series of representations with
the weights depending on one parameter, which can take arbitrary values.
In the case $so(2m+1)$ the representations admitted by the conditions 
form a discrete set. 

For the quadratic evaluation we have the analogous complete
characterization of the representations in the case where the 
$\e$-antisymmetric parts of the second weight components vanish, 
$\bar \lambda^{[2]}_i = 0$. The conditions can be written
in terms of three factors, two linear in the weights and one bilinear.       
In the cases $sp(2m)$ and $so(2m)$ there are series of representations with
the weights depending on up to three parameters, which can take arbitrary values.
In the case $so(2m+1)$ the representations admitted by the conditions 
have no more than one continuous parameter.

We have presented  examples of $L$ operators constructed explicitly
on the basis of an underlying algebra of Clifford or Heisenberg types
and an example related to the spinorial $R$ operator. 
We have compared the weights of the related representations 
to the ones obtained as solutions of the weight conditions for the linear
and quadratic evaluation.

\section*{Acknowledgments}
We thank A. Molev for useful discussions.

The work of D.K. was partially supported by 
the Armenian State Committee of Science grant 18T-132 and 
by the Regional Training Network on Theoretical Physics sponsored 
by Volkswagenstiftung Contract nr. 86 260.

%\section*{Appendix}

\end{document}